\pdfoutput=1
\documentclass[11pt,twoside,a4paper,cmspaper,final,collab]{cms-tdr}
\def\svnVersion{7fede73}\def\svnDate{2024/07/08}\def\cmsCernNoTag{CERN-EP-2024-038}\def\cmsCernDate{\today}\def\cmsMessage{Published in Physical Review D as \href{http://dx.doi.org/10.1103/PhysRevD.110.012002}{\doi{10.1103/PhysRevD.110.012002}.}}
\begin{document}\cmsNoteHeader{BPH-23-002}

\newcommand{\muon}      {\ensuremath{\PGm}\xspace}
\newcommand{\mup}       {\ensuremath{\PGmp}\xspace}
\newcommand{\mun}       {\ensuremath{\PGmm}\xspace}
\newcommand{\mumu}      {\ensuremath{\PGmp\PGmm}\xspace}
\newcommand{\pion}      {\ensuremath{\PGp}\xspace}
\newcommand{\piz}       {\ensuremath{\PGpz}\xspace}
\newcommand{\pip}       {\ensuremath{\PGpp}\xspace}
\newcommand{\pim}       {\ensuremath{\PGpm}\xspace}
\newcommand{\pipi}      {\ensuremath{\pip\pim}\xspace}
\newcommand{\pipm}      {\ensuremath{\PGppm}\xspace}
\newcommand{\Kp}        {\ensuremath{\PKp}\xspace}
\newcommand{\Km}        {\ensuremath{\PKm}\xspace}
\newcommand{\KS}        {\ensuremath{\PKzS}\xspace}
\newcommand{\La}        {\ensuremath{\PGL}\xspace}
\newcommand{\Sz}        {\ensuremath{\PGSz}\xspace}
\newcommand{\Xim}       {\ensuremath{\PGXm}\xspace}
\newcommand{\jpsi}      {\ensuremath{\PJGy}\xspace}
\newcommand{\psit}      {\ensuremath{\PGyP{2S}}\xspace}
\newcommand{\qs}        {\ensuremath{\PQq\PQs}\xspace}
\newcommand{\pvert}     {\ensuremath{\text{PV}}\xspace}
\newcommand{\B}         {\ensuremath{\PB}\xspace}
\newcommand{\Bp}        {\ensuremath{\PBp}\xspace}
\newcommand{\Bd}        {\ensuremath{\PBz}\xspace}
\newcommand{\Lb}        {\ensuremath{\PGLzb}\xspace}
\newcommand{\Xb}        {\ensuremath{\PGXmb}\xspace}
\newcommand{\Xbz}       {\ensuremath{\PGXzb}\xspace}
\newcommand{\Xbns}      {\ensuremath{\PGXb}\xspace}
\DeclareRobustCommand{\PGXzbP}[1]{{\HepParticleResonanceFull{\PGX}{b}{}{#1}{}{0}}\Xspace}
\newcommand{\Xbs}{\PGXzbP{5945}}
\newcommand{\Xbst}      {\ensuremath{\PGXstzb}\xspace}
\newcommand{\Xbsns}     {\ensuremath{\PGXstb}\xspace}
\newcommand{\Xbsm}      {\ensuremath{\PGXstmb}\xspace}
\newcommand{\Xbpz}      {\ensuremath{\PGXprzb}\xspace}
\newcommand{\Xbpns}     {\ensuremath{\PGXprb}\xspace}
\newcommand{\Sbz}       {\ensuremath{\PGSzb}\xspace}
\newcommand{\JMM}       {\ensuremath{\jpsi\to\mumu}\xspace}
\newcommand{\PMM}       {\ensuremath{\psit\to\mumu}\xspace}
\newcommand{\PJPP}      {\ensuremath{\psit\to\jpsi\pipi}\xspace}
\newcommand{\XbJX}      {\ensuremath{\Xb\to\jpsi\Xim}\xspace}
\newcommand{\XbPX}      {\ensuremath{\Xb\to\psit\Xim}\xspace}
\newcommand{\XbPsiXi}   {\ensuremath{\Xb\to\PGy\Xim}\xspace}
\newcommand{\XbJLK}     {\ensuremath{\Xb\to\jpsi\La\Km}\xspace}
\newcommand{\XbJSK}     {\ensuremath{\Xb\to\jpsi\Sz\Km}\xspace}
\newcommand{\jpsiXim}   {\ensuremath{\jpsi\Xim}\xspace}
\newcommand{\psitXim}   {\ensuremath{\psit\Xim}\xspace}
\newcommand{\psiXim}    {\ensuremath{\PGy\Xim}\xspace}
\newcommand{\jpsiLaKm}  {\ensuremath{\jpsi\La\Km}\xspace}
\newcommand{\jpsiSzKm}  {\ensuremath{\jpsi\Sz\Km}\xspace}
\newcommand{\XbP}       {\ensuremath{\Xb\pip}\xspace}
\newcommand{\dmX}       {\ensuremath{\Delta M}\xspace}
\newcommand{\Xbsd}      {\ensuremath{\Xbs\to\XbP}\xspace}
\newcommand{\pp}        {\ensuremath{\Pp\Pp}\xspace}
\newcommand{\LaPi}      {\ensuremath{\La\pim}\xspace}
\newcommand{\LamPPi}    {\ensuremath{\La\to\Pp\pim}\xspace}
\newcommand{\SzLaG}     {\ensuremath{\Sz\to\La\PGg}\xspace}
\newcommand{\XiLamPi}   {\ensuremath{\Xim\to\LaPi}\xspace}
\DeclareRobustCommand{\BcPrime}{\HepParticleResonanceFull{\PB}{c}{}{2S}{}{+}\xspace}
\DeclareRobustCommand{\BcPrimeStar}{\HepParticleResonanceFull{\PB}{c}{\ast{}}{2S}{}{+}\xspace}
\newcommand{\mXbfit}    {\ensuremath{m_{\Xb}^{\text{fit}}}\xspace}
\newcommand{\dmXfit}    {\ensuremath{\dmX^{\text{fit}}}\xspace}
\newcommand{\chisq}     {\ensuremath{\chi^2}\xspace}
\newcommand{\ips}       {\ensuremath{d_{xy}/\sigma_{d_{xy}}}\xspace}

\newcommand{\br}        {\ensuremath{\mathcal{B}}\xspace}
\newcommand{\ebran}     {\ensuremath{\,(\mathcal{B})}\xspace}
\newcommand{\sigef}     {\ensuremath{\sigma_{\text{eff}}}\xspace}

\ifthenelse{\boolean{cms@external}}{\providecommand{\cmsLeft}{upper\xspace}}{\providecommand{\cmsLeft}{left\xspace}}
\ifthenelse{\boolean{cms@external}}{\providecommand{\cmsRight}{lower\xspace}}{\providecommand{\cmsRight}{right\xspace}}
\newlength\cmsFigWidth\ifthenelse{\boolean{cms@external}}{\setlength\cmsFigWidth{0.48\textwidth}}{\setlength\cmsFigWidth{0.68\textwidth}}
\newlength\cmsTabSkip\setlength{\cmsTabSkip}{1ex}
\newlength\cmsTabSkipp\setlength{\cmsTabSkipp}{2ex}

\cmsNoteHeader{BPH-23-002}
\title{Observation of the \texorpdfstring{\XbPX}{Xib- to psi(2S) Xi-} decay and studies of the \texorpdfstring{\Xbs}{Xib(5945)0} baryon in proton-proton collisions at \texorpdfstring{$\sqrt{s} = 13\TeV$}{sqrt(s) = 13 TeV}}

\author*[cern]{The CMS Collaboration}

\date{\today}

\abstract{
The first observation of the decay \XbPX and measurement of the branching ratio of \XbPX to \XbJX are presented. The \jpsi and \psit mesons are reconstructed using their dimuon decay modes. The results are based on proton-proton colliding beam data from the LHC collected by the CMS experiment at $\sqrt{s} = 13\TeV$ in 2016--2018, corresponding to an integrated luminosity of 140\fbinv. The branching fraction ratio is measured to be $\br (\XbPX) / \br (\XbJX) = 0.84\,^{+0.21}_{-0.19} \stat \pm 0.10 \syst \pm 0.02 \,(\br)$, where the last uncertainty comes from the uncertainties in the branching fractions of the charmonium states. New measurements of the \Xbs baryon mass and natural width are also presented, using the \XbP final state, where the \Xb baryon is reconstructed through the decays \jpsiXim, \psitXim, \jpsiLaKm, and \jpsiSzKm. Finally, the fraction of \Xb baryons produced from \Xbs decays is determined.
}

\hypersetup{%
pdfauthor={CMS Collaboration},%
pdftitle={Observation of the Xib- to psi(2S) Xi- decay and studies of the Xib(5945)0 baryon in proton-proton collisions at sqrt(s) = 13 TeV},%
pdfsubject={CMS},%
pdfkeywords={CMS,  beauty baryon, heavy baryon, flavour spectroscopy}}

\maketitle 

\section{Introduction
\label{sec:intro}}

The \Xbns family consists of baryons that form isodoublets composed of a triplet of \PQb, \PQs, and \PQq quarks,
where \PQq corresponds to a \PQu or \PQd quark for the \Xbz and \Xb states,
respectively. Three such isodoublets that are neither orbitally nor radially excited
should exist~\cite{Th:diquark}. These include one with the spin of the light diquark
$j_{\qs} = 0$ and spin-parity of the baryon $J^P = 1/2^+$ (\Xbns ground states), 
one with $j_{\qs} = 1$ and $J^P = 1/2^+$ (\Xbpns),
and another with $j_{\qs} = 1$ and $J^P = 3/2^+$ (\Xbsns). 
The ground states were discovered more than a decade ago at the Fermilab Tevatron~\cite{XibD0,XibCDF, Xib0CDF} via 
their decays to \jpsiXim and $\PGXpc\pim$.
Three of the four states with $j_{\qs} = 1$ have been observed during the last decade
at the CERN LHC~\cite{BPH12001, LHCb_XibStar0, LHCb_XibPrime-Star-} via their
\XbP and $\Xbz\pim$ decays, as expected from theoretical 
predictions~\cite{Th:b_bar_1/Nc, Th:b_baryons, Th:m_bb_rel_quark_model}. 
The fourth state, \Xbpz, is expected to have a mass lower than the \XbP mass threshold,
making a strong decay to the \Xb baryon kinematically impossible. 
Several other more massive \Xbns resonances were also observed recently by the CMS and LHCb
Collaborations~\cite{LHCb_Xib6227, LHCb_Xib6227_2020, BPH20004, LHCb_Xib_1D, LHCb_Xbss}
via their decays to $\Xbz\pim$, \XbP, $\Xb\pipi$, $\Xbz\pipi$, $\Lb\Km$, and $\Lb\Km\pip$.
Various theoretical models and calculations predict a spectrum 
of excited \Xbns baryons~\cite{Th:m_bb_rel_quark_model, Th:b_bar_1/Nc, Th:b_baryons, Th:bb_in_quark_model, Th:regge_bb_rel, Th:faddeev_bb_spectr, Th:LQ_XQ_heavy_quark, Th:ITEP_Trusov, Th:sum_rule_Pwave, Th:12_32_heavy, Th:S_P_waves_single_bb, Th:heavy_w_chiral, Th:bb_hyperfine, Th:regge_traj, Th:spectr_123_bb}, and the observed resonances are considered to be $1P$ isodoublets of \Xbns or \Xbpns states, and a $1D$ doublet. 
However, larger data samples are needed to measure the quantum numbers of these resonances. 
There is also the possibility that some of the observed wide resonances could instead be 
unresolved overlapping narrow states.

Besides the searches for excited \Xbns states, the LHCb Collaboration 
has observed new ground-state \Xbns decays and determined some of 
their branching fractions~\cite{LHCb_Xb_Dph_Lch, LHCb_Xb_Lbpi, LHCb_Xbdec_pKK, LHCb_JpsiLamK, LHCb_Xb_charmless, LHCb_Xb_JpsiV0}. 
The spectrum of excited \Xbns baryons can be classified relatively easily, 
especially with the guidance of the similar and well-established \PGXc baryons~\cite{PDG}.
By contrast, the wide variety of decay modes available in the weak decay
of the ground-state baryons presents a significant theoretical challenge,
and predictions of the branching fractions to various final states are less straightforward. 
Multibody decays of \Xbns baryons can contain rich resonant
structures, including both conventional and exotic resonances,
such as excited \Xim states and the $P^\La_{\PGy\PQs}(4459)^0$ pentaquark
reported in the \XbJLK decay~\cite{LHCb_JpsiLamK_2020}. 
The search for new \Xbns decays is also important in the quest
for observing possible $CP$ violation~\cite{LHCb_Xbdec_pKK_CP}. 
In general, both weak decays of heavy baryons and their strongly decaying excitations 
can be described in the framework of heavy-quark effective theory 
(HQET)~\cite{Th:HQET, Th:hq_spectr_sym, Th:weak_decay_itep, Th:weak_decay_heavy, Th:hqet_nonper}.
Measurements of the decays and properties of both ground and excited \Xbns states provide coherent and 
complementary input to HQET, which could improve our understanding
of the quantum chromodynamic (QCD) mechanisms responsible for quark dynamics and the formation of hadrons.

In this paper, we study \Xb and \Xbs (also referred to as \Xbst here) baryon states using a sample of proton-proton ($\Pp\Pp$)
collisions from the LHC, collected by the CMS experiment in 2016--2018 at $\sqrt{s}=13\TeV$,
corresponding to an integrated luminosity of 140\fbinv~\cite{Sirunyan:2759951,CMS-PAS-LUM-17-004,CMS-PAS-LUM-18-002}. 
The inclusion of charge-conjugate states is implied throughout this paper, unless otherwise noted.
We report the first observation of the \XbPX decay and the measurement of its 
branching fraction with respect to the well-known \XbJX decay. 
In both signal and normalization channels, the charmonium states are reconstructed through
their dimuon decay modes, and \Xim decays to \LaPi with the following \LamPPi are used. 
Thus, the relative branching ratio $R$ is measured using the following expression: 
\begin{equation}\begin{aligned}
\ifthenelse{\boolean{cms@external}}
{ 
    R &= \dfrac{\br (\XbPX) }{\br (\XbJX) } = 
    \dfrac{N(\XbPX) }{N(\XbJX)} \\
    &\times \dfrac{\epsilon(\XbJX) }{\epsilon(\XbPX) }\, \dfrac{\br (\JMM)}{\br(\PMM)},
    \label{eq:R}
} 
{ 
    R = \dfrac{\br (\XbPX) }{\br (\XbJX) } = 
    \dfrac{N(\XbPX) }{N(\XbJX)} \dfrac{\epsilon(\XbJX) }{\epsilon(\XbPX) }\, \dfrac{\br (\JMM)}{\br(\PMM)},
    \label{eq:R}
} 
\end{aligned}\end{equation}
where $N$ and $\epsilon$ represent the measured number of signal events in data 
and the total efficiency from Monte Carlo (MC) simulation, respectively,
for each of the respective decay modes. The values of the branching fractions \br in the last term
are taken from the PDG~\cite{PDG}. 
Even though the value of $\br(\XbJX)$ is not known, the choice of this normalization channel
is quite natural since it has the same topology and similar kinematic properties as the signal
channel, reducing the systematic uncertainty in the ratio 
related to the reconstruction of the muons and the other charged particle tracks from the \Xb decays.

\begin{figure}[bh]
    \begin{center}
        \includegraphics[width=\cmsFigWidth]{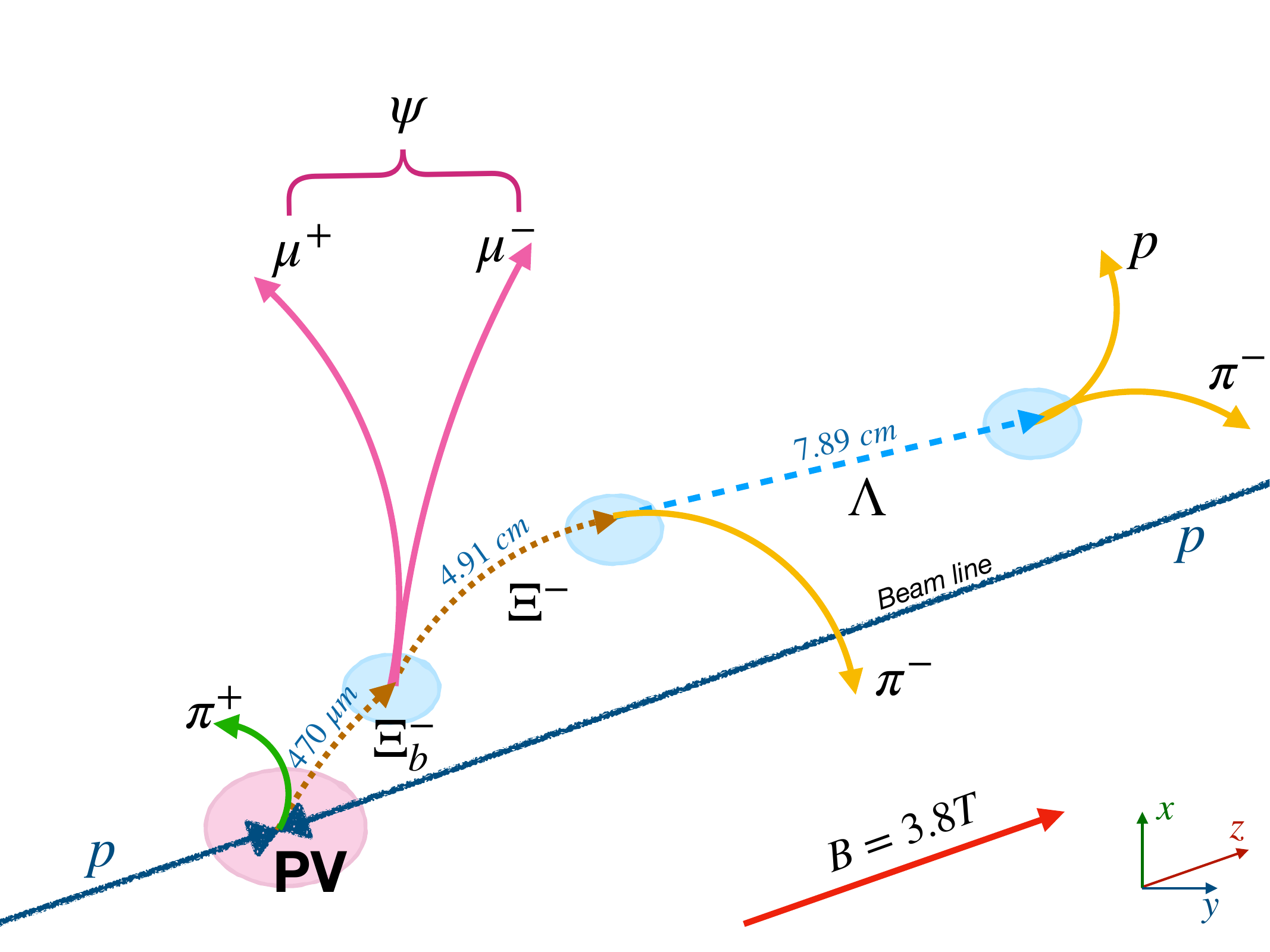}
        \includegraphics[width=\cmsFigWidth]{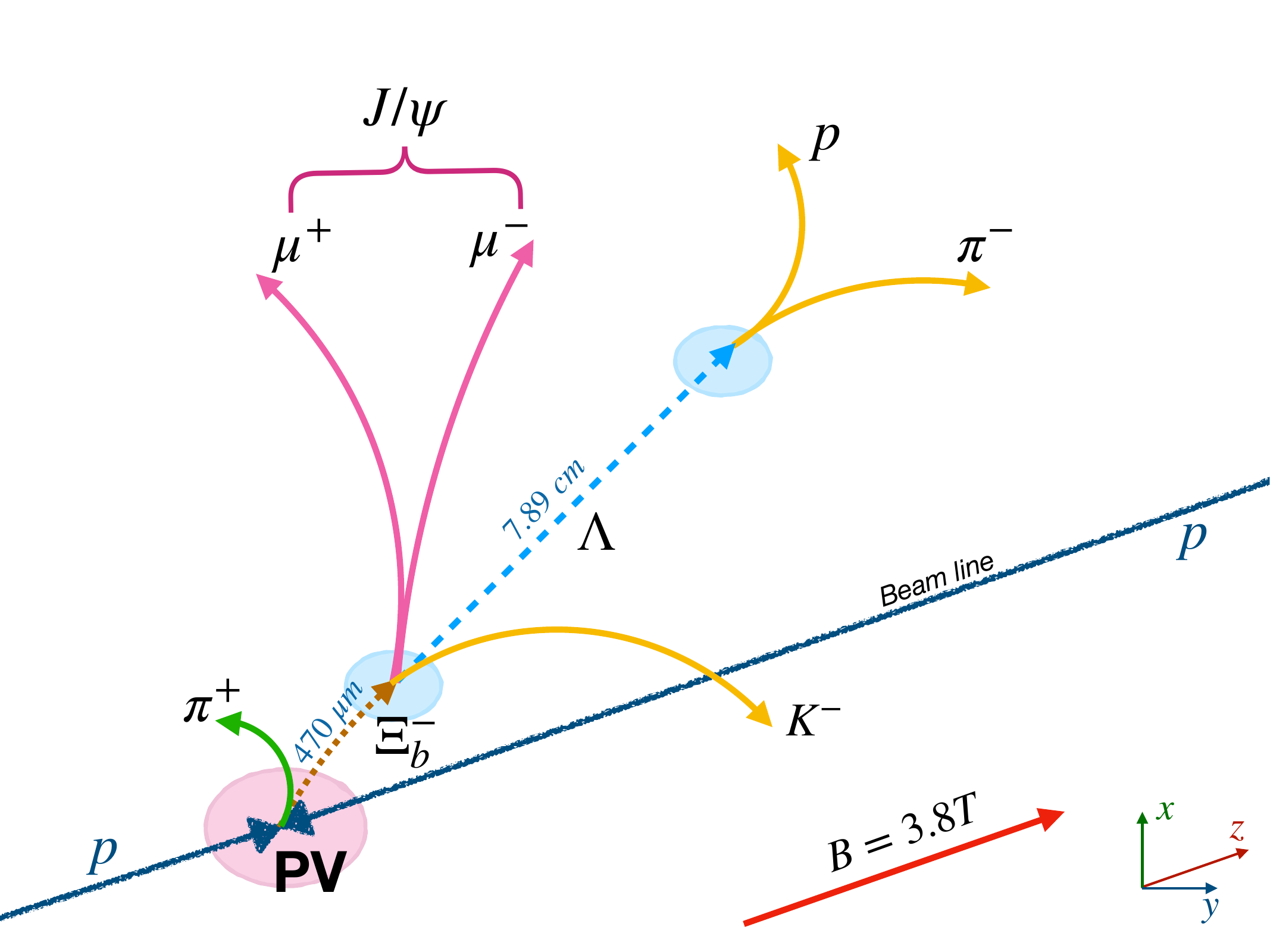}
    \end{center}
    \caption{
        The \Xbsd decay topology in which the \Xb baryon decays to \psiXim with $\PGy\to\mumu$ (upper) or
        \jpsiLaKm (lower), where \PGy refers to the \jpsi and \psit mesons.
        The distances given are the average decay lengths, $c\tau$.
        \label{fig:decay_topo}}
\end{figure}
We also determine the \Xbs baryon mass and natural width, using the \Xbsd decay. 
The ground-state \Xb is reconstructed via its decays to \jpsiXim, \psitXim, and \jpsiLaKm.
For the \XbPX decay, both \PJPP, with \JMM, and \PMM modes are used in the analysis, 
and \Xim is again reconstructed via the \LaPi channel.
For the \XbJLK decay, the presence of the partially reconstructed mode \jpsiSzKm,
where the low-energy photon from the \SzLaG decay is undetected, is 
included in the fit to the \XbP invariant mass spectrum.
Pictorial representations of the decay topologies for \Xbsd are shown in Fig.~\ref{fig:decay_topo}.

We also measure the ratio of the production cross sections $R_{\Xbs}$ for \Xbs and \Xb using the expression:
\ifthenelse{\boolean{cms@external}}{
\begin{multline}
    R_{\Xbs}\\
    \begin{aligned}
     =& \dfrac{\sigma(\pp\to\Xbs X) \, \br(\Xbsd)}{\sigma(\pp\to\Xb X)} \\ 
     =& \dfrac{N(\Xbsd)}{N(\Xb)} \dfrac{\epsilon(\Xb)}{\epsilon(\Xbsd)},
    \end{aligned}
    \label{eq:production}
\end{multline}
}{
\begin{equation}\begin{aligned}
    R_{\Xbs} & = \dfrac{\sigma(\pp\to\Xbs X) \, \br(\Xbsd)}{\sigma(\pp\to\Xb X)} \\ 
     & = \dfrac{N(\Xbsd)}{N(\Xb)} \dfrac{\epsilon(\Xb)}{\epsilon(\Xbsd)},
    \label{eq:production}
\end{aligned}\end{equation}    
}
where $N$ and $\epsilon$ refer to similar quantities as those in Eq.~\eqref{eq:R}.
Following an analogous CMS measurement of the \BcPrime and \BcPrimeStar 
production cross section ratios~\cite{BPH19001}, 
the \Xb baryon is reconstructed in the phase space region defined
by the \Xb baryon transverse momentum $\pt > 15$\GeV and rapidity $\abs{y} < 2.4$;
however, this measured ratio is intended to be representative of the entire phase space, 
given the small mass difference between the \Xbs and \Xb particles.
The \Xbs baryon was the first new particle observed by the CMS Collaboration,
using 5\fbinv of data from 2011~\cite{BPH12001}. 
With this paper we significantly improve and enrich our previous results for this state.
Tabulated results are provided in the HEPData record for this analysis~\cite{hepdata}.

\section{The CMS detector and simulated event samples
\label{sec:detector}}

The central feature of the CMS apparatus is a superconducting solenoid of 6\unit{m} 
internal diameter, providing a magnetic field of 3.8\unit{T}. Within the solenoid volume 
are a silicon pixel and strip tracker, a lead tungstate crystal electromagnetic 
calorimeter, and a brass and scintillator hadron calorimeter, each composed of a barrel 
and two endcap sections. Forward calorimeters extend the pseudorapidity ($\eta$) coverage provided 
by the barrel and endcap detectors. Muons are measured in gas-ionization detectors 
embedded in the steel flux-return yoke outside the solenoid. A more detailed description 
of the CMS detector, together with a definition of the coordinate system used and the 
relevant kinematic variables, can be found in Ref.~\cite{CMS}.
More recent changes to the detector are described in Ref.~\cite{CMS_Run3}

Muons are measured in the range $\abs{\eta} < 2.4$, with detection planes 
made using three technologies: drift tubes, cathode strip chambers, and resistive plate 
chambers. 
The single-muon trigger efficiency exceeds 90\% over the full $\eta$ range, 
whereas the efficiency to reconstruct and identify muons is greater than 96\%. 
Matching muons identified in the muon system to tracks measured in the silicon tracker
results in a relative \pt resolution for muons with \pt up to 100\GeV
of 1\% in the barrel and 3\% in the endcaps~\cite{CMS_muons}.
The silicon tracker used in 2016 measured charged particles within the range $\abs{\eta} < 2.5$. 
For nonisolated particles of $1 < \pt < 10\GeV$ and $\abs{\eta} < 1.4$, the track 
resolutions were typically 1.5\% in \pt and 25--90 (45--150)\mum in the transverse 
(longitudinal) impact parameter~\cite{CMS_tracks}. At the start of 2017, a new pixel 
detector was installed~\cite{Phase1Pixel}; the upgraded tracker measured particles 
up to $\abs{\eta} < 3$ with typical resolutions of 1.5\% in \pt and 20--75\mum in 
the transverse impact parameter~\cite{DP-2020-049} for nonisolated particles of $1 < \pt < 10\GeV$.
The default track selection used in CMS analyses is the ``high-purity'' requirement. 
Because low-momentum and displaced tracks share some features 
with nongenuine tracks such as not pointing back to the \pp collision vertex
and having fewer measurement points, 
the high-purity selection is less efficient for these tracks 
and so the less-restrictive ``loose'' requirement is often used.

Events of interest are selected using a two-tiered trigger system~\cite{Khachatryan:2016bia}. 
The first level (L1), composed of custom hardware processors, uses information from the 
calorimeters and muon detectors to select events at a rate of around 100\unit{kHz} within a fixed latency 
of about 4\mus~\cite{Sirunyan:2020zal}. The second level, known as the high-level trigger 
(HLT), consists of a farm of computing processors running a version of the full event 
reconstruction software optimized for fast processing, and reduces the event rate to 
around 1\unit{kHz} before data storage.
The events used in this analysis were selected at L1 by requiring 
the presence of at least two muons, and at the HLT by requiring 
that the two muons have opposite sign (OS), with various $\abs{\eta}$ and \pt thresholds,
compatible with being produced in the dimuon decay of \jpsi or \psit mesons
by requiring corresponding invariant mass windows.

The \PYTHIA~8.240 package~\cite{Pythia8} with the CP5 underlying event tune~\cite{CMS:2019csb}
is used to simulate the production of the \Xb and \Xbs states (where the \Sbz baryon, 
with a modified mass value, is used as a proxy for the \Xbs state). 
The \EVTGEN~1.6.0~\cite{EvtGen} program models the decays 
\Xbsd, \XbJX, \XbPX, \XbJLK, and \XbJSK with a phase-space decay model,
followed by the \PMM, \PJPP, and \JMM decays.
Final-state photon radiation is included using \PHOTOS~3.61~\cite{PHOTOS,PHOTOS_2}.
The generated MC events are then passed to a detailed \GEANTfour-based
simulation~\cite{GEANT4} of the CMS detector,
which includes the  long-lived hyperon decays \XiLamPi and \LamPPi.
The simulated events are then put through the same trigger
and reconstruction algorithms used for the collision data.
The simulation includes effects from multiple \pp interactions in the same
or nearby bunch crossings (pileup) with a multiplicity distribution
matching that in data.

\section{Event reconstruction and selection
\label{sec:evesel}}

The \Xb ground state is reconstructed using two main decay modes: \XbPsiXi (followed 
by $\PGy\to\mumu$), where \PGy refers to the \jpsi and \psit mesons, or \XbJLK. 
We also reconstruct the decay chain \XbPX, \PJPP to increase the number of events
for the \XbP studies.  
In all the cases, the \jpsi meson is identified through its dimuon decay.
The selection criteria, described below, are mainly 
inherited from Ref.~\cite{BPH20004}. 

The reconstruction chain requires two OS muons forming a good-quality vertex, passing the CMS soft-muon
selection~\cite{CMS_muons}, and with each having $\pt(\muon) > 3\GeV$ and $\abs{\eta(\muon)}<2.4$.
To be of good quality, the fit to a dimuon common vertex must have a 
\chisq vertex fit probability greater than 1\%. 
These requirements reinforce those applied at the trigger level during the online data taking.
A \jpsi or \psit candidate is required to have a dimuon invariant mass within 100\MeV 
of the corresponding world-average mass~\cite{PDG}, which is about 3 times the invariant mass resolution.
Further, a kinematic constraint to the known \PGy meson mass~\cite{PDG}
is applied to the selected dimuon candidates.

The \La candidates are formed from displaced two-prong vertices, assuming the decay \LamPPi,
as described in Ref.~\cite{CMS_tracks_V0}.  
The higher-momentum track is associated with the proton and the lower-momentum track
with the pion.
A \La candidate must have $\pt > 1.8\GeV$, and the $\Pp\pim$ invariant mass must be within 
10\MeV of the known \La mass~\cite{PDG} after the tracks are refit to a common vertex, 
corresponding to about 3 times the invariant mass resolution.
The vertex fit is then repeated with the $\Pp\pim$ invariant mass constrained
to the \La mass, and its momentum recomputed. 
The \chisq probability of this fit must be greater than 1\%.

For the \XbPsiXi channel, the \Xim candidates are reconstructed by combining 
each selected \La candidate with a charged particle track, assumed to be a pion.
The track must have $\pt > 0.3\GeV$ 
and satisfy the loose requirement~\cite{CMS_tracks}. 
A kinematic vertex fit of the \XiLamPi decay is performed, 
and the \chisq probability is required to be greater than 1\%.  
The \LaPi invariant mass must be within 10\MeV
of the known \Xim mass~\cite{PDG}, which is about 3 times the invariant mass resolution.  
The resulting \Xim candidate must have $\pt > 2.5\GeV$.
Because \La particles mainly decay much further from the \Xim decay vertex than our vertex
resolution, we set a requirement
on the pointing angle $\alpha(\La, \Xim)$ between
the momentum of the \La candidate and the vector from the \Xim decay vertex
to the \La decay vertex in the plane perpendicular to the beam direction (the transverse plane) of $\cos\alpha(\La, \Xim) > 0.99$.

{\tolerance=800
To reconstruct the decay chain \XbPX, \PJPP, two additional OS tracks passing the 
high-purity requirement~\cite{CMS_tracks} are assigned the charged pion mass and added to the process.  
The higher-momentum pion must have $\pt > 0.6\GeV$, and
the other pion $\pt > 0.35\GeV$.
The invariant mass of the \psit candidate, calculated via the formula 
$M(\mumu\pipi) - M(\mumu) + m^{\mathrm{PDG}}(\jpsi)$, is required to be  within 18\MeV
of the known \psit mass~\cite{PDG}, corresponding to about 3 times the invariant mass resolution.
Using this variable removes the \JMM detector invariant mass resolution
from the measurement of the $M(\mumu\pipi)$ invariant mass.
Here, and throughout the paper, the symbol $M$ represents a reconstructed invariant mass
and $m^{\mathrm{PDG}}$ the PDG world-average mass~\cite{PDG}.
\par}

The \Xb candidates are selected by using the \mup, \mun, and \Xim particles
in a kinematic fit that constrains their momentum vectors to a common vertex
and the dimuon invariant mass to the world-average \jpsi or \psit mass~\cite{PDG}. 
For the decay chain \XbPX, \PJPP, 
the two additional pions described above are added to the \Xb vertex fit.  
From all the reconstructed \pp collision vertices in an event, the primary vertex (\pvert)
is chosen as the one with the smallest pointing angle.
The pointing angle is the angle between the \Xb candidate momentum and the
vector joining the \pvert with the reconstructed \Xb candidate decay vertex.
If any of the tracks used in the
\Xb candidate reconstruction are included in the fit of the chosen 
\pvert, they are removed, and the \pvert is refit. 
The selected \Xb candidates are required to have 
$\pt(\Xb) > 10\GeV$ and a \chisq vertex fit probability greater than 1\%.
The pion from the \XiLamPi decay must satisfy an impact parameter significance requirement 
$\ips > 1$, where $d_{xy}$ is the closest distance between the track and the chosen \pvert 
in the transverse plane, and $\sigma_{d_{xy}}$ is its uncertainty.
For the decay chain \XbPX, \PJPP, we require that the two pion tracks each have $\ips > 0.4$.
The pointing angle $\alpha$ between the \Xim momentum and the vector from the \Xb decay vertex 
to the \Xim vertex in the transverse plane must satisfy $\cos{\alpha(\Xim, \Xb)} > 0.999$.
The analogous angle between the \Xb momentum and the vector
from the \pvert to the \Xb vertex is required to have $\cos{\alpha(\Xb, \pvert)} > 0.99$.
Additionally, the distance $L_{xy}$ between the PV and the \Xb decay vertex in 
the transverse plane must fulfill the requirement $L_{xy} / \sigma_{L_{xy}} > 3$,
where $\sigma_{L_{xy}}$ is its uncertainty.

For the \XbJLK decay channel, the \jpsi and \La candidates
are reconstructed in the same way as described above, with the additional requirement $\pt(\La) > 2\GeV$. 
However, instead of adding a pion track to the subsequent \XiLamPi fit, 
a charged particle track with a kaon mass assignment is selected. 
The track must have $\pt > 1.4\GeV$ 
and satisfy the high-purity requirement~\cite{CMS_tracks}. 
The \Xb candidates are obtained by performing a kinematic vertex fit to the 
\mup, \mun, \La, and \Km candidates, along with the same \jpsi mass constraint and PV selection
as for the \XbPsiXi channel. The kaon impact parameter significance must satisfy $\ips > 0.5$
with respect to the chosen PV. 
Because of the higher background in this channel
more restrictive kinematic and topological requirements are applied:
$\pt(\Xb) > 15\GeV$ and $\cos{\alpha(\Xb, \pvert)} > 0.999$,
along with the same requirements as above on the vertex fit and $L_{xy} / \sigma_{L_{xy}}$.
For both the \psiXim and \jpsiLaKm channels, multiple \Xb candidates, found in the same event
(which happens in 1--4\% of events depending on the channel) are not discarded.

Since the lifetime of the excited \Xbns states is expected to be negligible, 
the \XbP candidates are formed by combining the selected \Xb candidates with each
charged particle track originating from the PV and satisfying 
the loose requirement~\cite{CMS_tracks} as done in Ref.~\cite{BPH18007}, 
which are given the charged pion mass.
The pion charge must be opposite to that of the pion from \XiLamPi or the kaon from \XbJLK.
The mass difference variable $\dmX = M(\XbP) - M(\Xb) - m^{\mathrm{PDG}}_\pip$
is used instead of $M(\XbP)$ since it is characterized by a better invariant mass resolution since the the effect of the \Xb invariant mass resolution is removed.
From simulation studies, this variable is found to be insensitive to potential invariant mass shifts
caused by the missing low-energy photon from the \SzLaG decay.
As developed in Ref.~\cite{BPH19003}, the \Xb candidate and all the
tracks forming the PV are refit to a common vertex, 
further improving the \XbP invariant mass resolution
from $1.07\pm0.07$ to $0.74\pm0.04\MeV$ (statistical uncertainties only), 
as determined from simulation studies.
If multiple \XbP candidates (where the multiplicity comes from the soft pion reconstruction) 
in an event pass the selection requirements
(which happens in 10--15\% of events depending on the \Xb channel),
only the highest \pt candidate is kept, 
which is found from simulation studies to improve the signal purity.
The multiplicity estimation and a single-candidate selection 
is performed in the signal region of $[0,\,0.06]\GeV$ of the \dmX variable.

\section{Observation of the \texorpdfstring{\XbPX}{Xib- to psi(2S) Xi-} decay and studies of the \texorpdfstring{\Xb}{Xib-} signal
    \label{sec:Xb_signals}}

\begin{figure*}[ht]
    \centering   
    \includegraphics[width=0.48\textwidth]{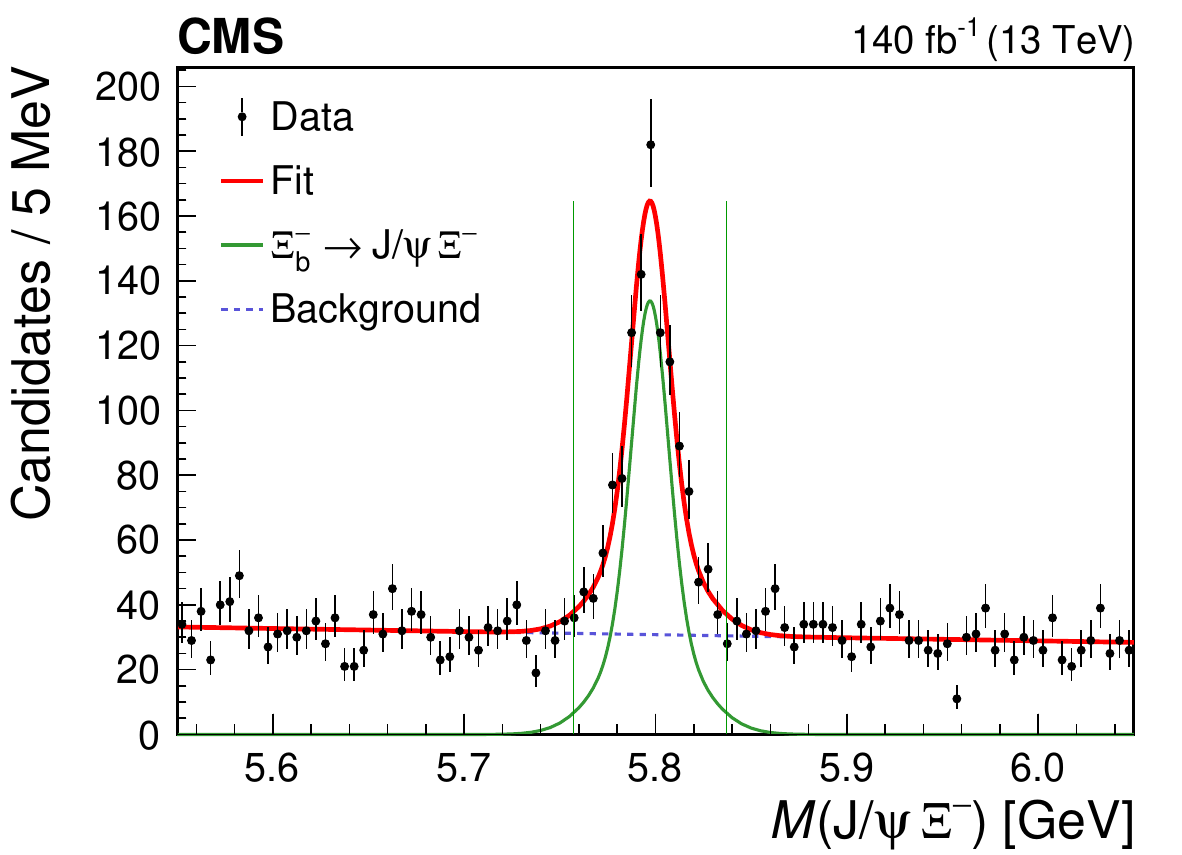}
    \includegraphics[width=0.48\textwidth]{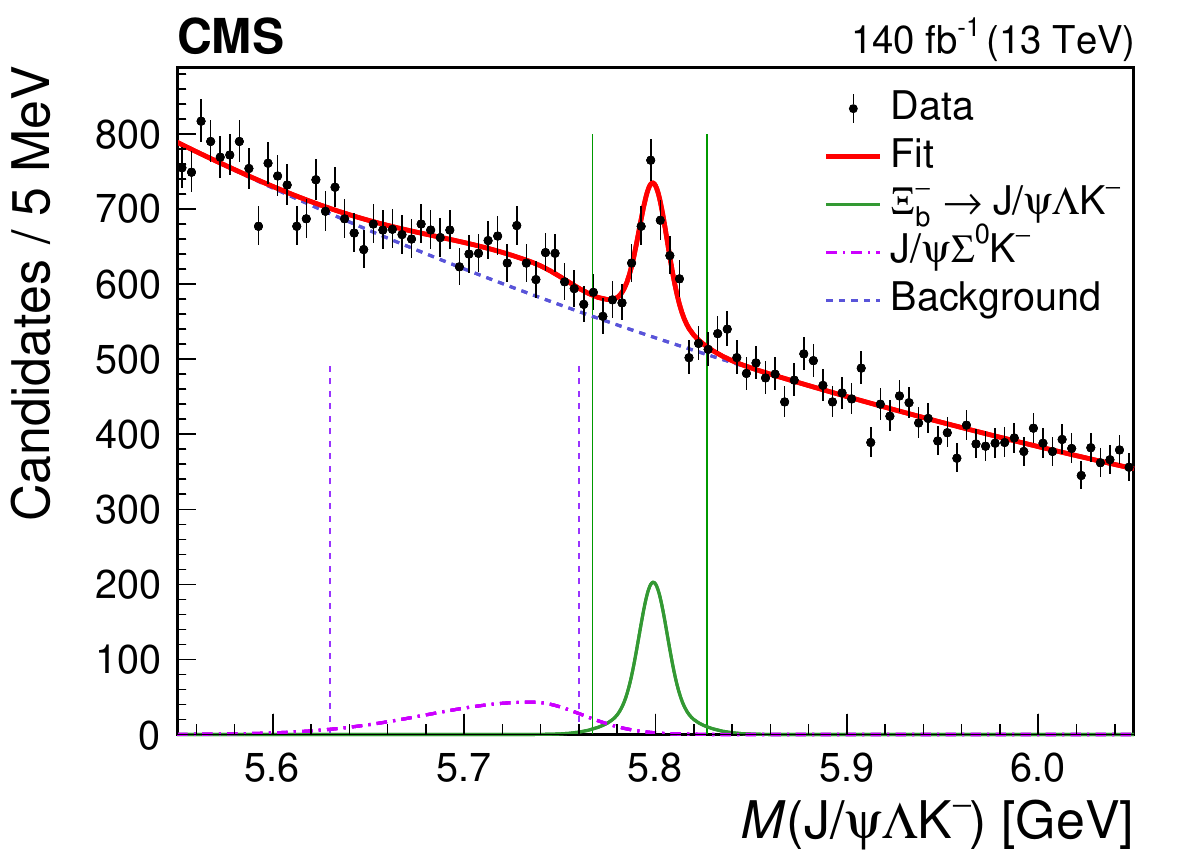}
    \includegraphics[width=0.48\textwidth]{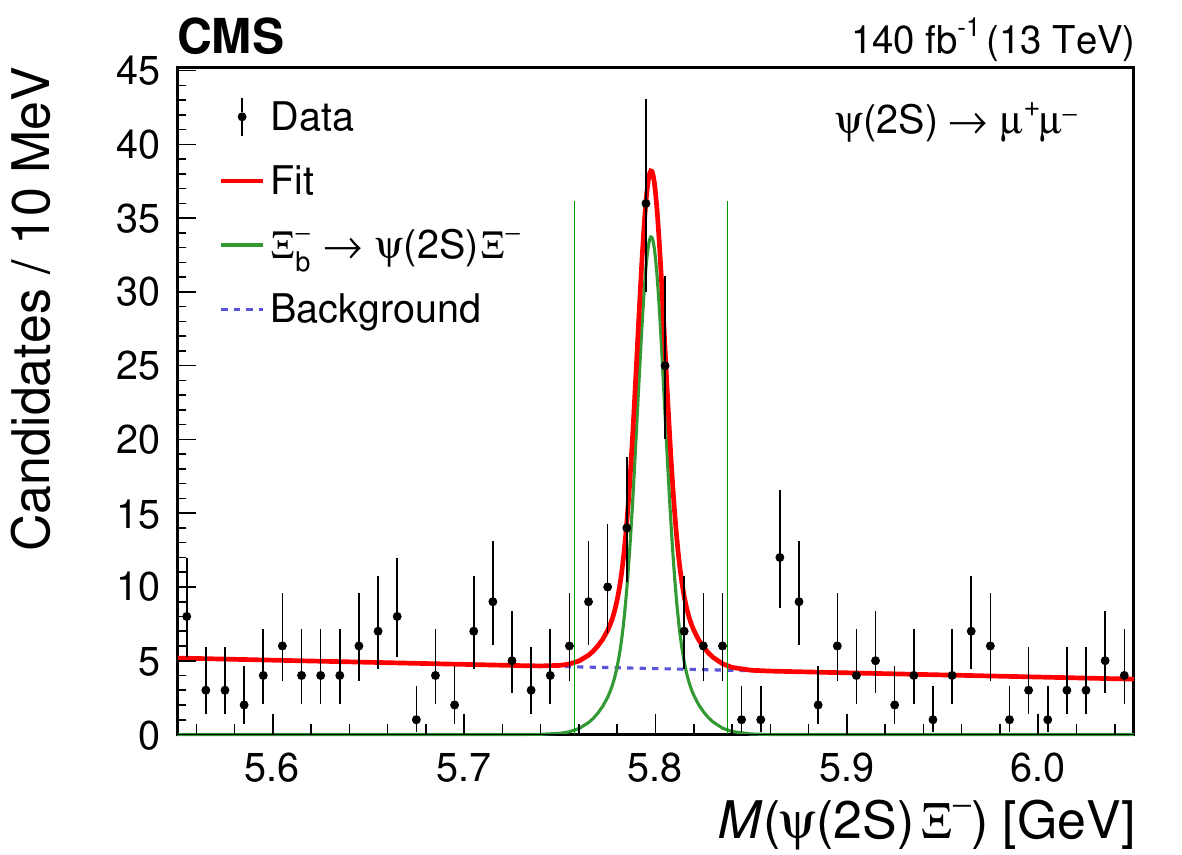}
    \includegraphics[width=0.48\textwidth]{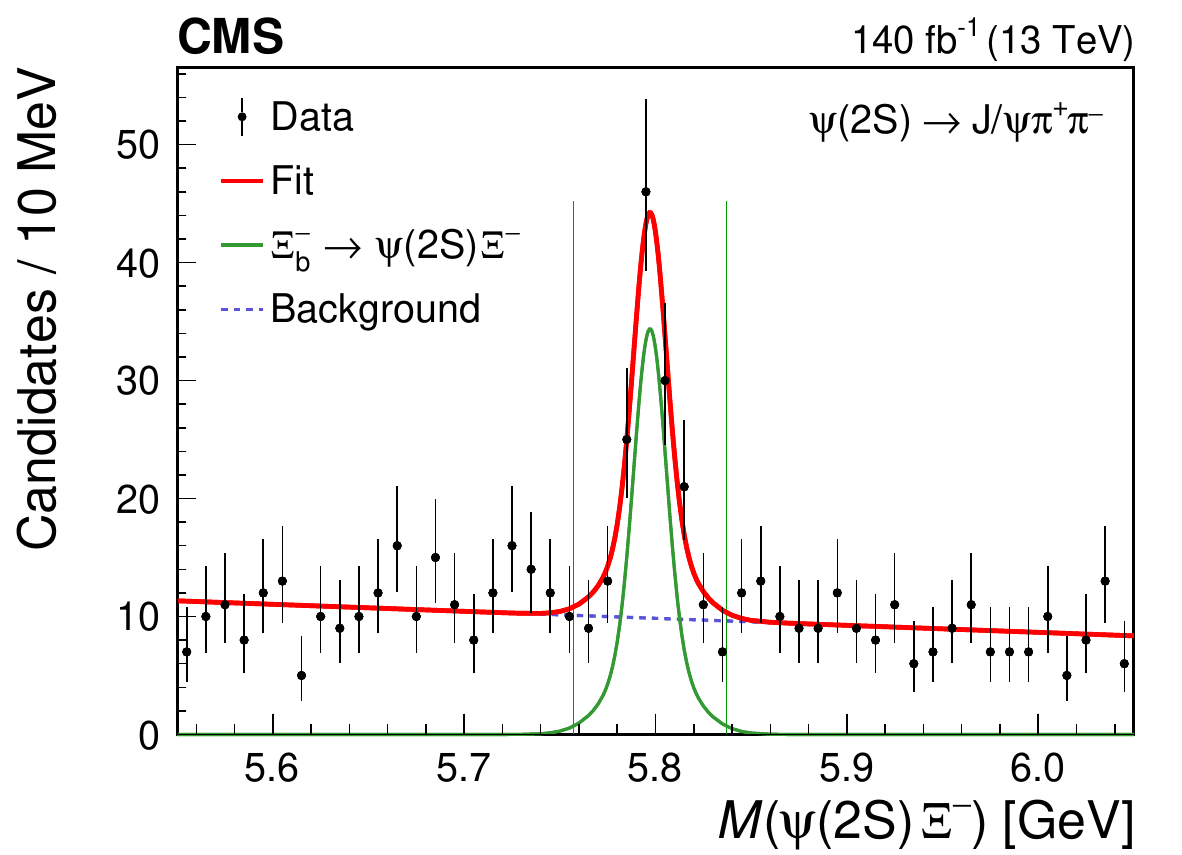}    
    \caption{
        Invariant mass distributions of the selected \jpsiXim (upper left), \jpsiLaKm (upper right), 
        and \psitXim [lower row, with \PMM (left) and \PJPP (right)] candidates.
        The data are shown by the points, while the vertical bars represent the statistical uncertainties.  
        The overall fit result is shown by the solid red curve, with the signal 
        and background contributions given by the solid green
        and dashed blue curves, respectively. 
        The vertical lines around each peak display the mass window required for a \Xb candidate to be used in the \Xbs studies. 
        The dotted-dashed curve in the upper right plot shows the fitted contribution
        from the \XbJSK decay, with the accompanying vertical dotted lines
        indicating the mass window for this mode.
        \label{fig:Xb_signal}
    } 
\end{figure*}

The invariant mass distributions of the selected \jpsiXim, \jpsiLaKm, and \psitXim 
(with both \PMM and \PJPP) candidates are shown in Fig.~\ref{fig:Xb_signal}. 
An unbinned extended maximum likelihood fit is performed on each of these distributions.
For all four channels, the signal component is described using the sum of two
Gaussian functions with a common mean, whose widths and ratio between them 
are fixed to those determined from MC simulation. 
However, both widths are allowed to scale by the same free parameter in the fit to give a 
better description of the data.
The background is described with a first-order polynomial for the \jpsiXim and \psitXim channels,
and an exponential function for the \jpsiLaKm. 
In the latter fit, the signal contribution from the partially reconstructed \XbJSK decays 
is taken into account by including an asymmetric Gaussian (also known as a skew normal) function
in the fit, whose shape parameters are fixed to those found from simulation studies.

\begin{table*}[bht]
    \centering
    \topcaption{
        The number of signal events $N$, the mean \Xb mass 
        \mXbfit, and the effective \Xb width \sigef from the fits
        to the \Xb invariant mass distributions for each of the \Xb decay 
        channels. The uncertainties are statistical only.
    }
    \begin{scotch}{lccc} 
        Decay channel
        &   $N$
        &   \mXbfit (\MeVns)
        &   \sigef (\MeVns)
        \\[\cmsTabSkipp]
        \hline
        \XbJX
        &   $846 \pm 40\phantom{1}$
        &   $5797.1 \pm 0.6$
        &   $16.3 \pm 1.0$
        \\
        \XbJLK
        &   $920 \pm 98\phantom{1}$
        &   $5798.8 \pm 0.9$
        &   $11.9 \pm 1.5$
        \\
        \XbJSK
        &   $880 \pm 170$
        &   \NA
        &   \NA
        \\
        \XbPX (with \PMM)
        &   $\phantom{8}74 \pm 11\phantom{0}$
        &   $5797.7 \pm 1.4$
        &   $11.1 \pm 2.0$
        \\
        \XbPX (with \PJPP)
        &   $\phantom{8}90 \pm 14\phantom{0}$
        &   $5797.2 \pm 1.7$
        &   $13.1 \pm 2.8$
        \\
    \end{scotch}
    \label{tab:Xb_signal}
\end{table*}

The number of signal events $N$, the mean reconstructed \Xb invariant mass \mXbfit, and the effective \Xb width
\sigef from the fit are given in Table~\ref{tab:Xb_signal}
for each of the \Xb decay channels, along with their statistical uncertainties.
The value of \sigef is calculated as $\sqrt{\smash[b]{f_1\sigma_1^2 + (1-f_1)\sigma_2^2}}$,
where $\sigma_1$ ($\sigma_2$) is the width of the first (second) Gaussian function, 
and $f_1$ is the 
fraction of signal events from the fit associated with the first Gaussian function.
The measured resolution of the different channels is within the expectations from the 
available phase space and the final-state threshold proximity. 
The fitted \Xb mass values are consistent with the world-average value
$m_{\Xb}^{\mathrm{PDG}} = 5797.0\pm 0.6\MeV$~\cite{PDG}.

This is the first observation of the \XbPX decay.
Its local statistical significance is evaluated with the likelihood ratio technique,
comparing the likelihood value from a fit to a signal-plus-background hypothesis
to that for a background-only hypothesis. 
Since the conditions of Wilks' theorem~\cite{Wilks_LL_ratio} are satisfied, 
the asymptotic formulae of Ref.~\cite{Cowan_ssymptotic} 
[Eqs.~(12) and (52)] are used to determine the \XbPX signal significance, 
which is found to be well above 5 standard deviations for both the \PMM and \PJPP modes.

For the \XbP studies described in the next section, the \Xb candidates must have
an invariant mass within 40 (30)\MeV of the \mXbfit value for the \jpsiXim and \psitXim
(\jpsiLaKm) decay channels. This corresponds to about (2.5--3) times \sigef, 
as shown by the solid vertical lines around the peaks in Fig.~\ref{fig:Xb_signal}.
For the partially reconstructed \jpsiSzKm decay channel, 
a mass window of 5.63--5.76\GeV, as in Ref.~\cite{BPH20004} and
shown by the vertical dotted lines in Fig.~\ref{fig:Xb_signal} (upper right), is
used for the reconstructed \Xb mass.

\section{Studies of the \texorpdfstring{\Xbs}{Xib*0} baryon
    \label{sec:XbPStudy}}

\begin{figure*}[ht!]
    \centering
        \includegraphics[width=0.48\textwidth]{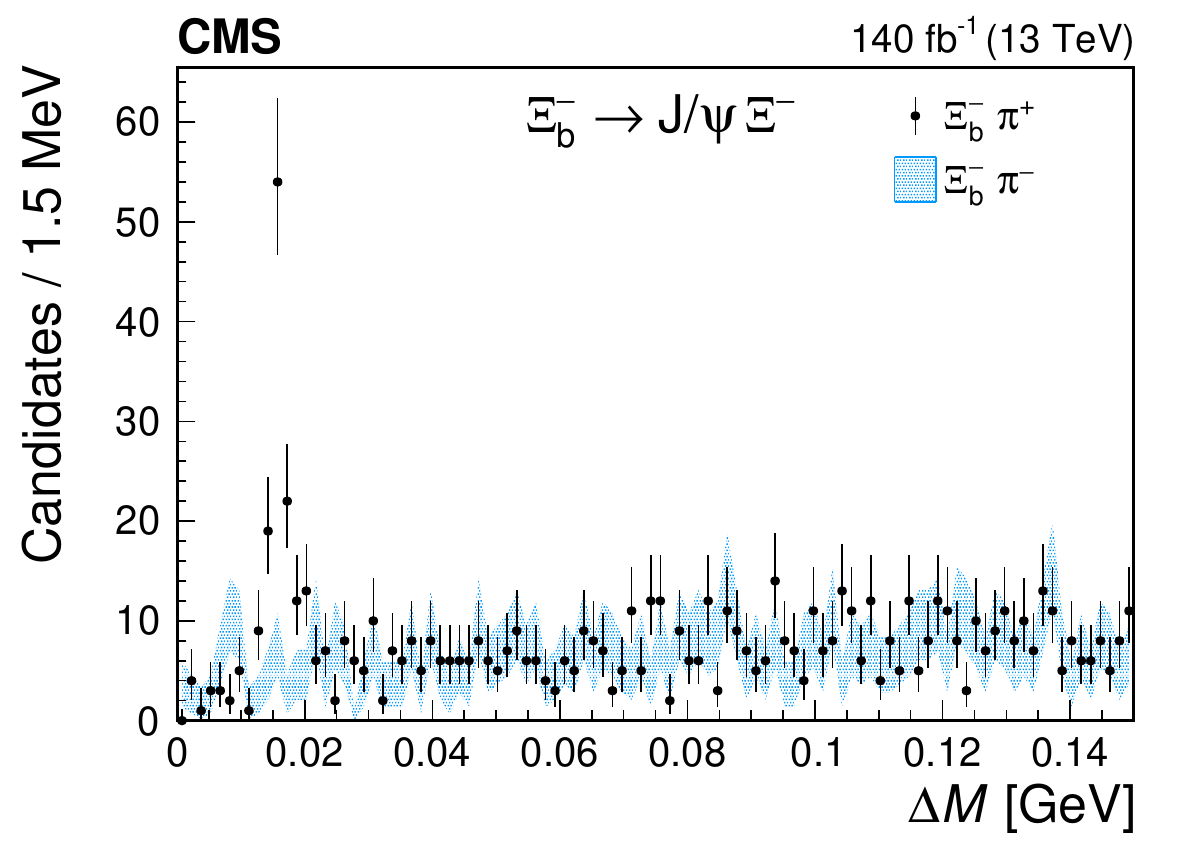}
        \includegraphics[width=0.48\textwidth]{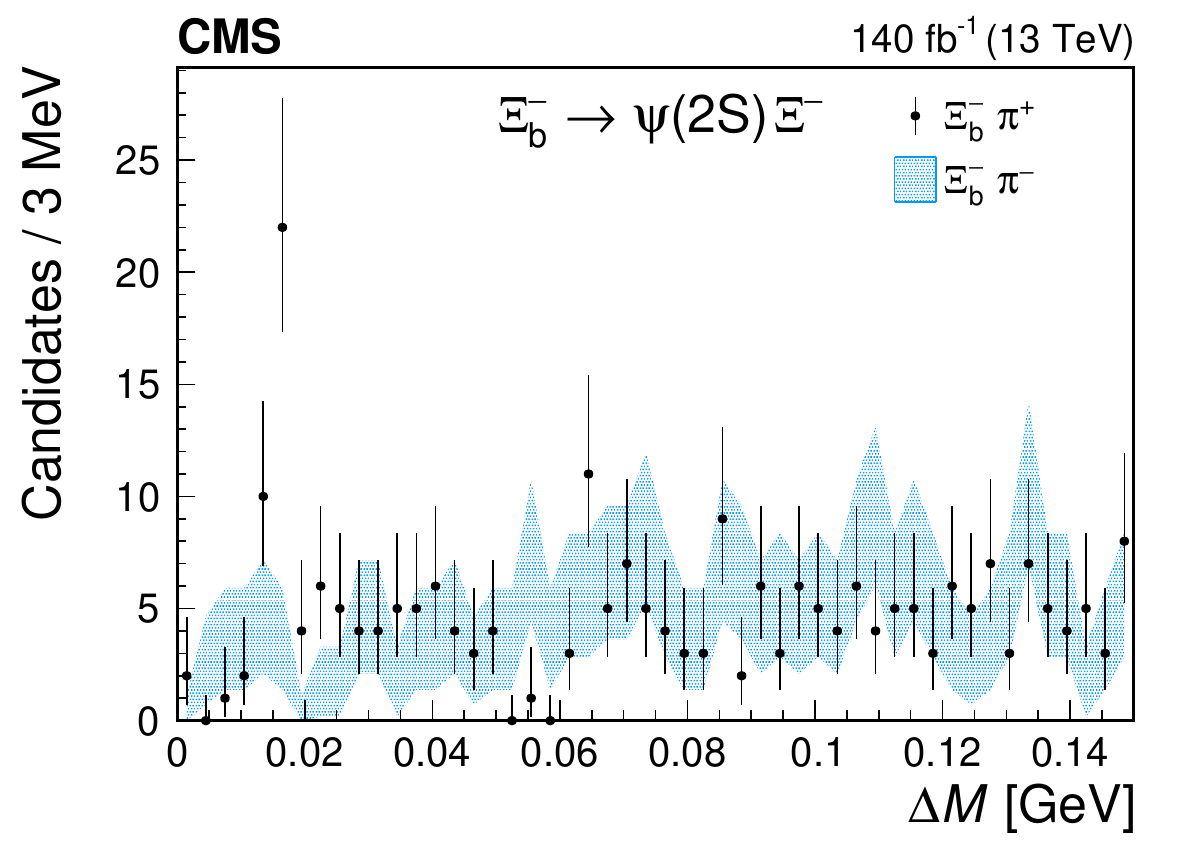}
        \includegraphics[width=0.48\textwidth]{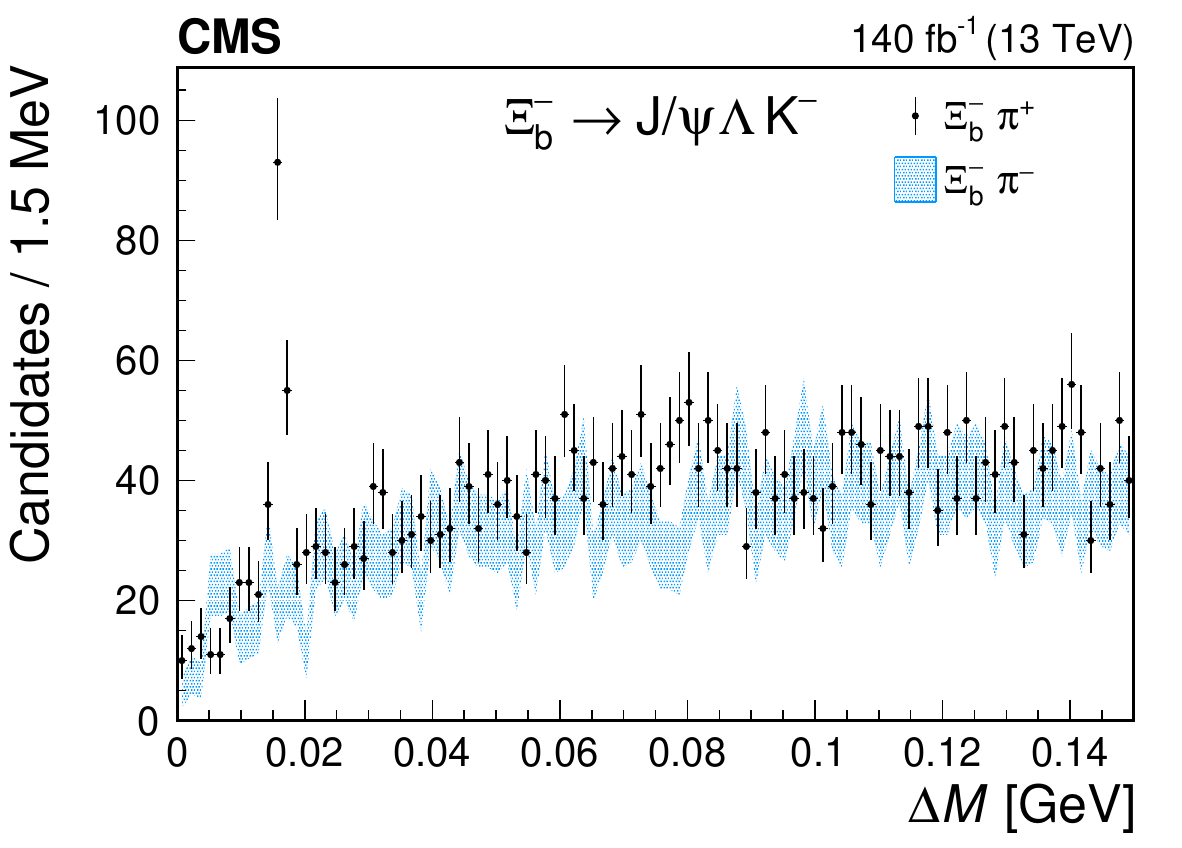}
        \includegraphics[width=0.48\textwidth]{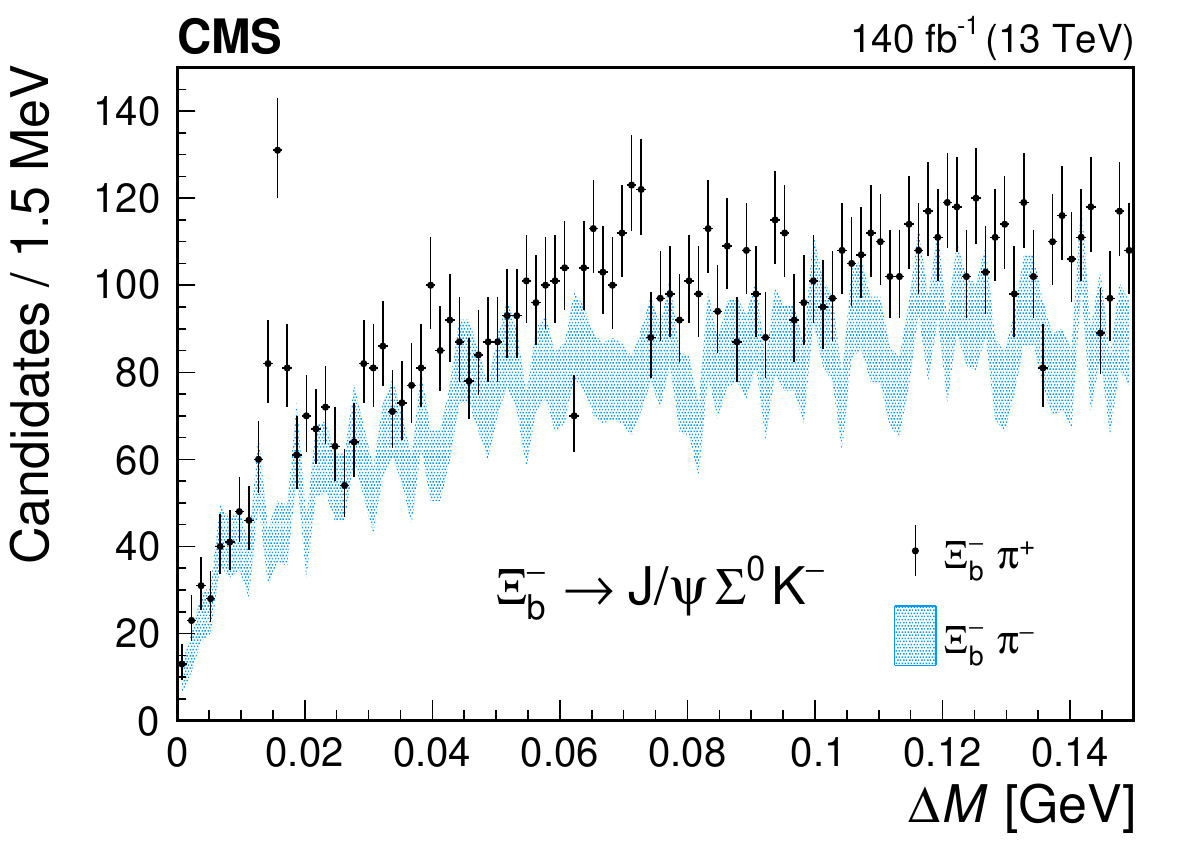}
    \caption{
        The mass difference \dmX distribution of the selected $\Xb\pipm$ candidates
        for the decay channel labeled on each plot.  
        The points show the correct-sign combinations and the blue bands the wrong-sign.
        The vertical bars on the points and the length of the bands
        represent the statistical uncertainties in each distribution, respectively.
        \label{fig:Xbs_nofit}
    } 
\end{figure*}

The measured \dmX distributions found by combining the selected \Xb candidates,
as defined in  Section~\ref{sec:Xb_signals}, with charged particle tracks, 
consistent with coming from the PV and assumed to be pions, are shown in Fig.~\ref{fig:Xbs_nofit}.
The distributions are shown separately for the \XbJX, \XbPX (combined \PJPP and \PMM modes),
\XbJLK, and \XbJSK channels. 
A significant near-threshold peak is evident in all 4 distributions, 
in agreement with previous CMS~\cite{BPH12001} and LHCb~\cite{LHCb_XibStar0, LHCb_Xbss} results.
The \dmX distribution for the same-sign $\Xb\pim$ control sample
is also displayed in Fig.~\ref{fig:Xbs_nofit}.
It shows no evidence of a peak and is consistent with the \XbP combinatorial background.
No other structures are observed in this \dmX region 
for either the \XbP or $\Xb\pim$ distributions.

We fit the \Xbs signal using a relativistic Breit--Wigner function,
which accounts for the non-negligible natural width $\Gamma(\Xbs)$, convolved with a Gaussian
function describing the invariant mass resolution,
whose parameters are extracted from MC simulation.
Lattice QCD calculations~\cite{Th:lattice} give $\Gamma(\Xbs)=0.51\pm0.16\MeV$, 
the ${ }^3 P_0$ model predicts 0.85\MeV~\cite{Th:3P0_model}, 
and the latest LHCb result finds $\Gamma(\Xbs) = 0.87\pm0.06\stat\pm0.05\syst\MeV$~\cite{LHCb_Xbss}.
The simulation studies predict that the invariant mass resolution is slightly different
for each \Xb baryon decay channel, except for the \XbJSK mode, where the missing low-energy photon
from the \Sz baryon decay produces a much wider peak with a 26\% larger mass resolution.
In all cases, the measured widths from the fully reconstructed decay modes are in 
agreement within their uncertainties.

\begin{figure*}[th!]
    \centering
    \includegraphics[width=0.48\textwidth]{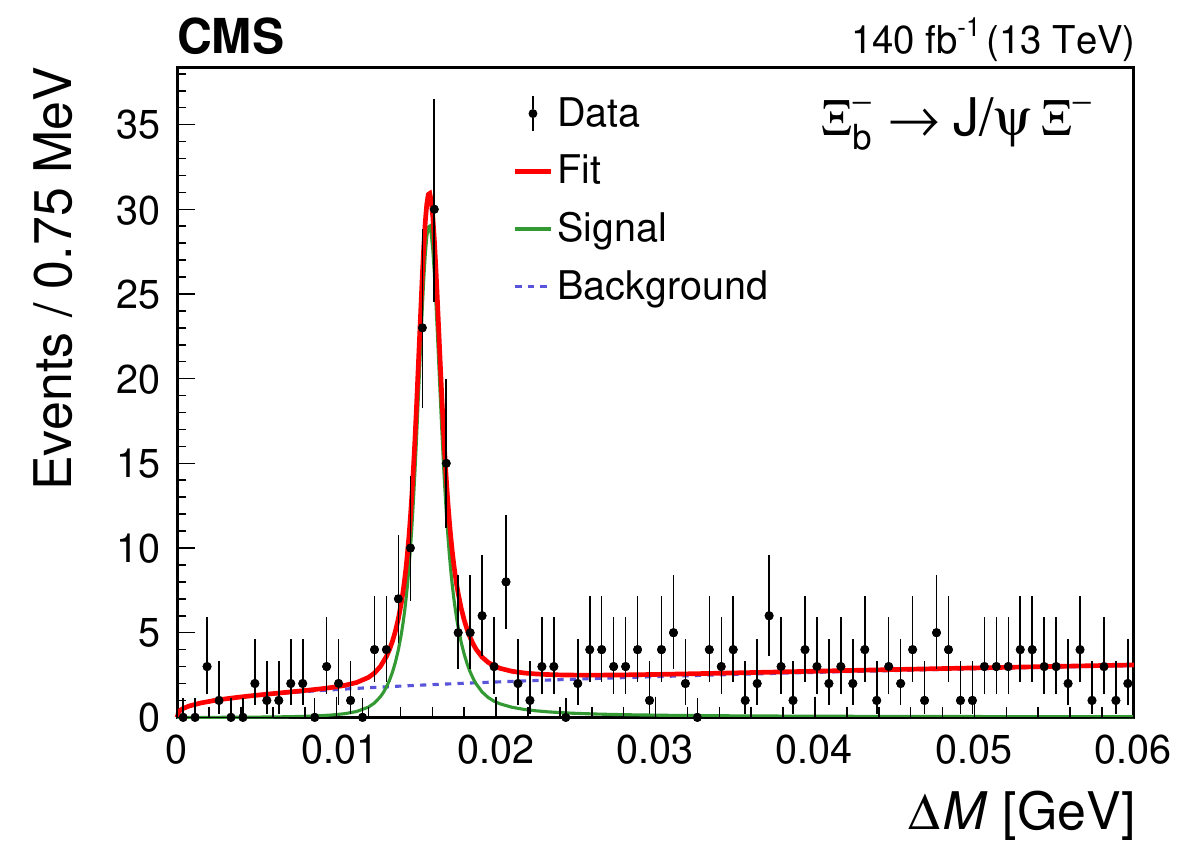}
    \includegraphics[width=0.48\textwidth]{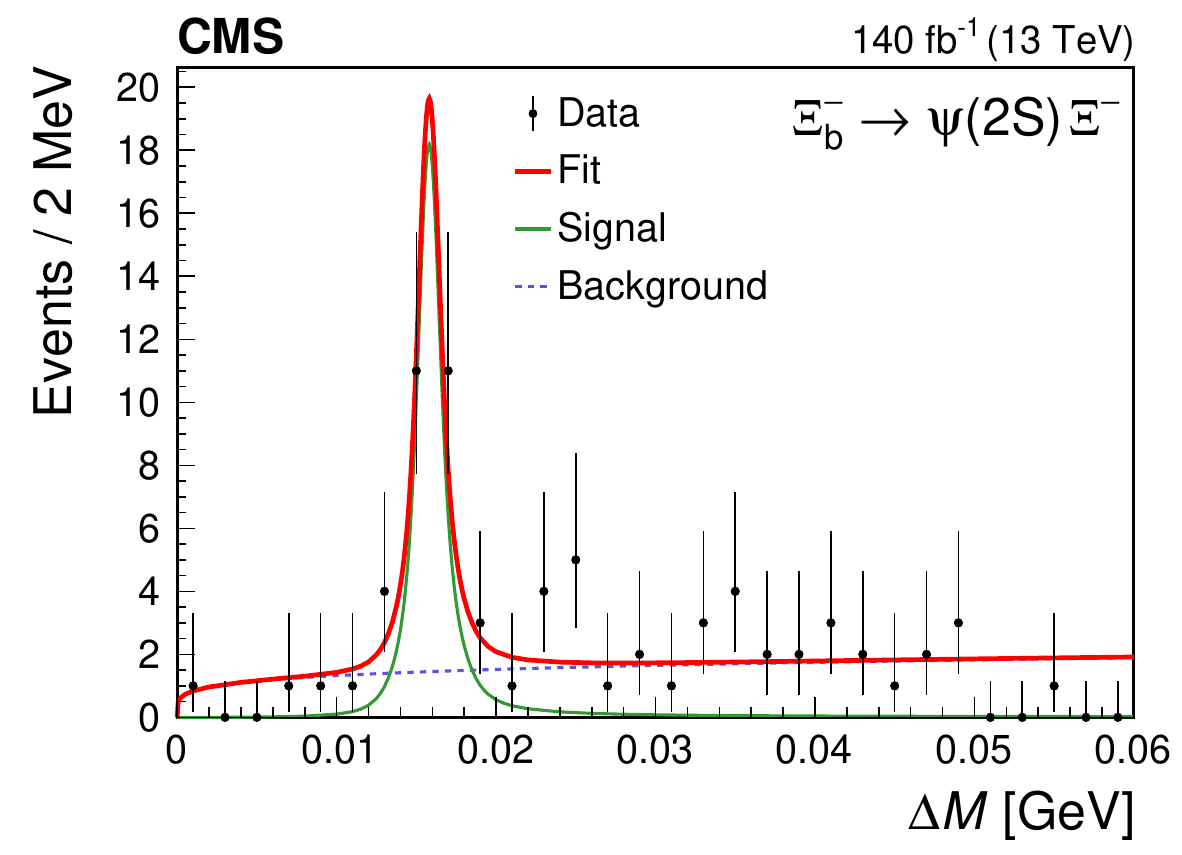}
    \includegraphics[width=0.48\textwidth]{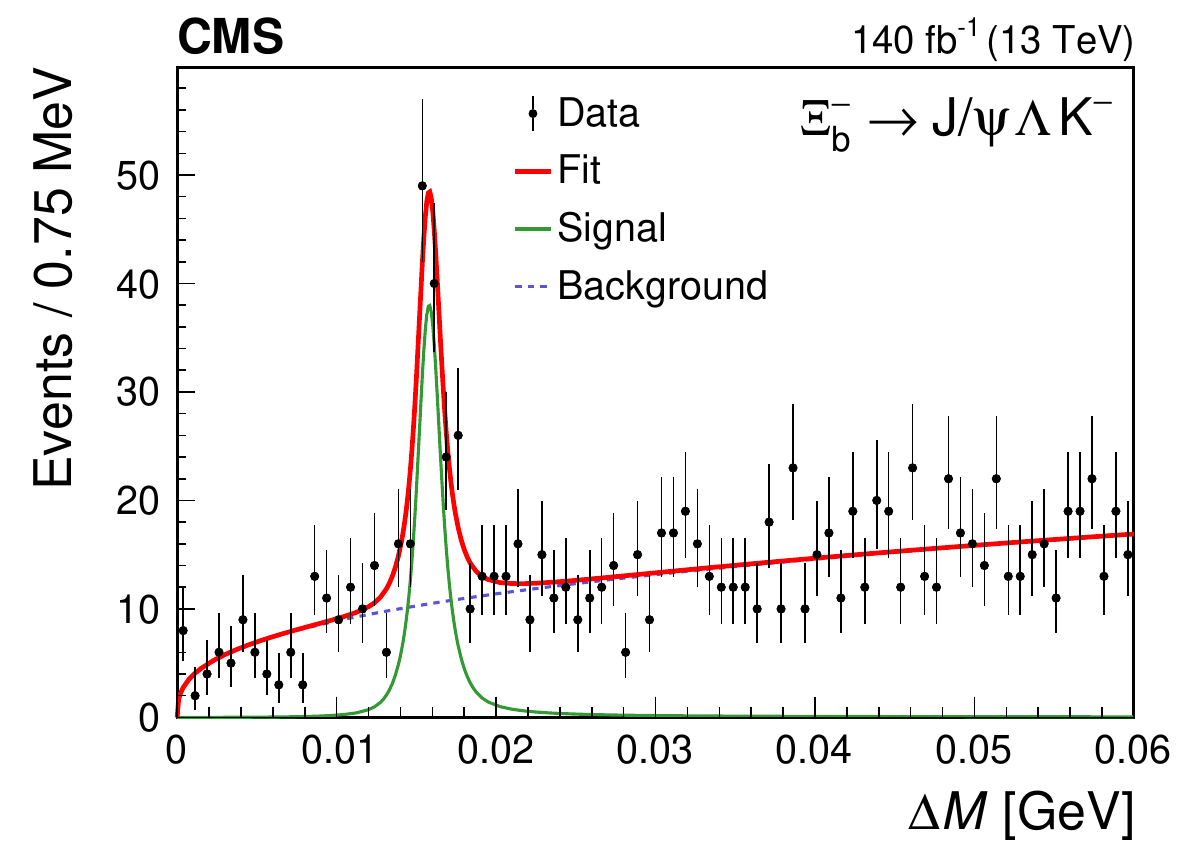}
    \includegraphics[width=0.48\textwidth]{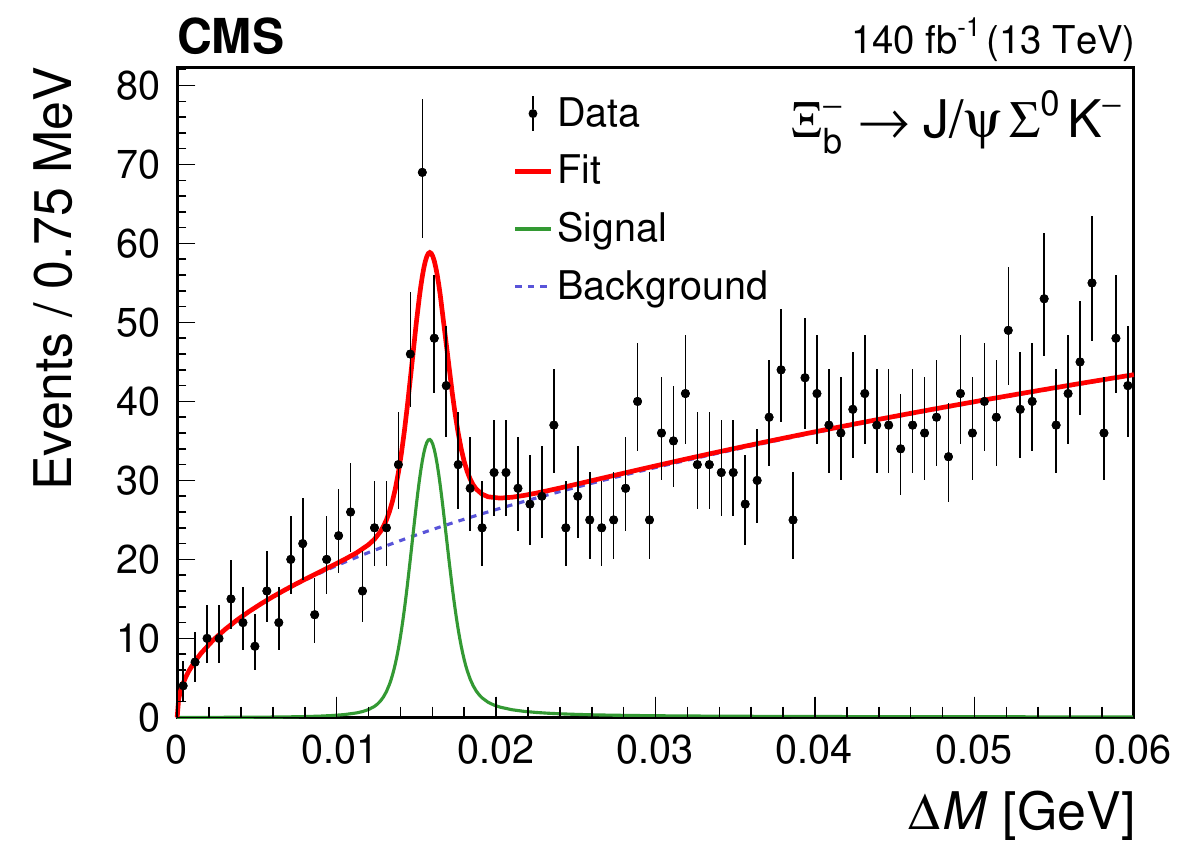}
    \caption{
        Results of the simultaneous fits to the \dmX invariant mass distributions 
        for the \Xbs candidates in the decay channels given in each plot.  
        The points show the data, with the vertical bars representing the statistical
        uncertainty. The solid red curve  displays the overall fit result, 
        with the solid green and dashed blue curves showing the signal
        and background contributions, respectively.
        \label{fig:Xbs_signal}
    } 
\end{figure*}

An unbinned extended maximum likelihood simultaneous fit of all four channels is applied, 
where the \Xbs mass and natural width are constrained to be equal for all the channels,
while the mass resolutions, yields, and background parameters are different. 
The background component is modeled with a threshold function $(\dmX)^\alpha$,
where $\alpha$ is a free parameter. 
The fit results are shown in Fig.~\ref{fig:Xbs_signal}, and the fitted signal yields 
are given in Table~\ref{tab:Xbs_signal}. 

The measured mass difference and natural width of the \Xbs state are 
$\dmXfit = 15.810 \pm 0.077\MeV$ 
and $\Gamma(\Xbs) = 0.87\,^{+0.22}_{-0.20}\MeV$, respectively, 
where the uncertainties are statistical only.

\begin{table}[bh!]
    \centering
    \topcaption{
        The fitted signal yields of the \Xbsd decay for each of the listed \Xb decay channels.
        Uncertainties are statistical only.
    } 
    \renewcommand{\arraystretch}{1.3}
    \begin{scotch}{lc} 
        Decay channel
            &   $N(\Xbs)$                               \\  \hline
    \XbJX   &   $\phantom{1}97\,^{+13}_{-12}$           \\
    \XbPX   &   $\phantom{1}24\,^{+6\phantom{2}}_{-5}$  \\
    \XbJLK  &   $124\,^{+17}_{-16}$                     \\
    \XbJSK  &   $155\,^{+22}_{-20}$
    \end{scotch}
    \renewcommand{\arraystretch}{1.0}
    \label{tab:Xbs_signal}
\end{table}

\section{Efficiency and production ratio measurements
\label{sec:eff}} 

While in general the analysis uses events collected by a combination of different 
dimuon HLT paths, for the measurements of the ratios of efficiencies and the resulting
branching fractions and production cross sections, 
a single dedicated trigger suitable for the decay topology is required in order to simplify 
the efficiency estimations and reduce the trigger-related systematic uncertainty. 
For the \XbJX and \XbPX channels, we use an inclusive dimuon HLT path, 
requiring the presence in the event of a \jpsi [\psit\negthinspace\negthinspace] meson 
with \pt exceeding 25~(18)~\GeV and decaying into two OS muons.
This HLT path is only used for the 2017--2018 sample,
while for the 2016 sample the similar trigger requires a minimum \pt of
20~(13)~\GeV for the \jpsi~[\psit\negthinspace\negthinspace] meson. 
In the case of the \XbJLK channel, we use an HLT path requiring the presence 
of a \JMM decay and an additional track consistent with originating from the dimuon vertex
and having $\ips>2$. The dimuon vertex must also be displaced from the PV, 
by requiring $L_{xy}/\sigma_{L_{xy}} > 3$.

These requirements are much stricter than those discussed in Section~\ref{sec:evesel} --- 
most \PGy from \Xb decays are populated within the 10--20\GeV range of \pt.
Thus, using them causes a significant decrease in the signal yields for the $\PGy\Xim$ channels.
Redoing the fitting procedure with the new requirements leads to total signal yields of 
$103^{+14}_{-13}$ and $38^{+8}_{-7}$ events for \XbJX and \XbPX (\PMM mode), respectively. 
The \XbJLK signal with the tighter HLT requirement results in $606^{+67}_{-64}$ events.
The fits to the \Xbsd \dmX distributions are performed separately for each of the decay channels,
with $\Gamma(\Xbs)$ fixed to the value found from the simultaneous fit.
The resulting signal yields are  $13 \pm 4$ and  $74 \pm 11$ events 
for the \jpsiXim and \jpsiLaKm decay modes, respectively. 

The efficiencies for the signal and normalization channels are calculated using simulated
MC samples of events that have passed the more-restrictive HLT paths described above.
The total efficiency includes several factorizable contributions such as the trigger,
detector acceptance, and decay channel reconstruction efficiencies.
The detector acceptance term is calculated as the ratio of the 
number of generator-level events within the CMS kinematic acceptance to the number of generated 
events without any restrictions (within the full phase space region).
Efficiencies for different years of data taking are estimated separately and then 
combined with weights corresponding to the integrated luminosity collected in each year. 

Since we measure branching fractions and production cross sections with respect to
normalization channels, only the ratios of such efficiencies are needed.
Thus, for example, the systematic uncertainties
associated with the muon, charged particle track, and \La candidate reconstruction are reduced.
Table~\ref{tab:eff} reports three efficiency ratios, 
where the first is used in measuring the quantity $R$, 
the ratio of branching fractions defined in Eq.~\eqref{eq:R}, 
and the latter two for finding the \Xbs / \Xb production cross section ratio
using two different decay channels: \jpsiXim and \jpsiLaKm.

\begin{table*}[bht]
    \centering
    \topcaption{
        The measured efficiency ratios and their statistical uncertainties.
    }
    \renewcommand{\arraystretch}{1.2}
    \begin{scotch}{cc} 
        Efficiency ratio
        &   Value
        \\  
        \hline
        \rule[0pt]{23pt}{0pt}$\epsilon(\XbJX) / \epsilon(\XbPX,\,\PMM)$
        &  $0.304 \pm 0.014$    \\
        $\epsilon(\XbJX) / \epsilon(\Xbsd,\,\XbJX)$
        &  $1.645 \pm 0.108$    \\
         $\epsilon(\XbJLK) / \epsilon(\Xbsd,\,\XbJLK)$
        &  $1.941 \pm 0.085$
    \end{scotch}
    \renewcommand{\arraystretch}{1.0}
    \label{tab:eff}
\end{table*}

Using the measured signal yields, the efficiency ratio, and Eq.~\eqref{eq:R},
we determine the ratio $R$ of the branching fraction for the 
newly observed \XbPX decay to that of the \XbJX decay to be
\begin{equation*} 
R = 0.84^{+0.21}_{-0.19},
\end{equation*} 
where the uncertainty is coming from the uncertainty in the measured yields. 
The uncertainty in the ratio of efficiencies is treated separately
as a systematic uncertainty, as described in Section~\ref{sec:syst_ratio}.

Applying Eq.~\eqref{eq:production}, the ratio $R_{\Xbs}$ of the \Xbs to \Xb 
production is separately measured using two \Xb decay 
channels: \jpsiXim and \jpsiLaKm. The results are 
\begin{equation*} 
    R_{\Xbs}^{\jpsiXim} = 0.21 \pm 0.04
\end{equation*}  
and
\begin{equation*} 
    R_{\Xbs}^{\jpsiLaKm} = 0.24 \pm 0.03,
\end{equation*} 
where the uncertainties are statistical only (again, the efficiency uncertainties are discussed in Section~\ref{sec:syst_ratio}).
These uncertainties account for the correlation between $N(\Xbs)$ and $N(\Xb)$, given that the former is a subset of the latter. 
Both values, obtained with fully independent data and simulation samples, 
are in good agreement with each other and with the previous measurement 
by the LHCb Collaboration~\cite{LHCb_XibStar0}.

\section{Systematic uncertainties 
    \label{sec:syst}}

The systematic uncertainties in the measurements given above are divided into two categories.
The first is related to the uncertainties in the measured efficiency ratios
and the \Xb and \Xbs signal yields.  
The second covers the uncertainties in the measured mass difference
and natural width of the \Xbs baryon.

\subsection{Systematic uncertainties in the measured ratios
\label{sec:syst_ratio}}

Many systematic uncertainties related to muon reconstruction and identification, 
trigger effects and efficiencies, and charged particle track and \La candidate reconstruction 
cancel out in the measured ratio $R$ due to the identical topologies of the \XbPX and \XbJX decays.
There is a similar cancellation in the determination of the production
cross section ratio $R_{\Xbs}$, where the only topological difference
between \Xbs and \Xb is an additional track from the \Xbsd decay. 

The systematic uncertainty related to the choice of fit functions used to describe the signal
and background shapes in the invariant mass fits is evaluated by varying the functions used
and recording the change in the number of signal events.
For the three \Xb decay channels, we first perform the fit with the resolution scaling parameter 
for the sum of the two Gaussian functions set to unity and note the change in the fit results.  
We then use a Student's t distribution~\cite{Studentt} to model the signal,
with the mean and width allowed to be free and the $n$ parameter 
(corresponding to the number of degrees of freedom) fixed from the simulation.  
This function, being symmetric and bell-shaped, also models a heavy-tailed distribution
and thus is found to be a reliable alternative to the sum of two Gaussian functions.
A single Gaussian function with free parameters is also tried for fitting the \psitXim and 
\jpsiXim signals. Using the largest change in the number of events, the resulting systematic
uncertainty in $R$ from this source is 8.8\%.

Two alternative background functions are considered in fitting the
\jpsiXim and \psitXim invariant mass distributions: an exponential
function and a second-order polynomial. For the more complicated
background shape in the \jpsiLaKm distribution, we switch from
an exponential function to a second-order polynomial. The resulting
systematic uncertainty in $R$ from this source is estimated as 4.5\%.
The combined signal-plus-background \Xb fit model uncertainties are estimated as
4.0 and 6.9\% in the $R_{\Xbs}^{\jpsiXim}$ and $R_{\Xbs}^{\jpsiLaKm}$ values, respectively.

The alternative functions used in fitting the \Xbsd \dmX distribution are described in the next
subsection when the systematic uncertainties in the measured \Xbs mass and width are 
discussed. The resulting systematic uncertainties due to the fitting
functions in the $R_{\Xbs}$ production cross section ratio are 7.7 and 
6.7\% for the \jpsiXim and \jpsiLaKm decay modes, respectively.

For the $R$ measurement, given that we are using different HLT paths for the \XbJX and the 
\XbPX signals, a cross-check of the correctness and robustness of such a procedure is performed.  
The similar branching fraction ratio $R_{\Bp} = \br(\Bp\to\psit\Kp) / \br (\Bp\to\jpsi\Kp)$ 
was measured with the triggers we use for the \Xb signals, 
and the resulting value of $0.601 \pm 0.030$ is consistent with the 
world-average value~\cite{PDG} $0.605 \pm 0.021$.
The 5\% precision of the $R_{\Bp}$ value is taken conservatively
as an additional systematic uncertainty in the $R$ measurement. 

As mentioned above, an additional source of uncertainty in the $R_{\Xbs}$ measurement
comes from identifying the extra pion in the \Xbsd decay. 
The uncertainty in the tracking reconstruction efficiency for the low-\pt pion
is estimated as 5.2\%~\cite{BPH18003}.

The uncertainty related to the finite size of the MC samples is also considered
as a systematic uncertainty. It is estimated from the statistical uncertainty in the 
determinations of the efficiency ratios from the MC simulation.  
This corresponds to a systematic uncertainty of 4.6\% in $R$, 
and 6.5 and 4.4\% in $R_{\Xbs}$ for the \jpsiXim and \jpsiLaKm modes, respectively.

The systematic uncertainties in the $R$ and $R_{\Xbs}$ measurements are summarized in
Tables~\ref{tab:syst_R} and \ref{tab:syst_R_Xbs}, respectively, along with the total
systematic uncertainties, calculated from the sum in quadrature of the
individual sources.

\begin{table}[ht]
    \centering
    \topcaption{
        Systematic uncertainties in percent in the ratio $R$ from the different sources 
        and the total uncertainty.
        }
        \begin{scotch}{lc}
            Source 
            & Uncertainty (\%)
            \\  
            \hline
            Signal model
            &   8.8
            \\  
            Background model
            &   4.5
            \\  
            $R_{\Bp}$ uncertainty
            &	5.0
            \\
            MC finite size
            &   4.6
            \\[\cmsTabSkip]
            Total
            &   12.0
        \end{scotch}
        \label{tab:syst_R}
\end{table}

\begin{table}[ht]
    \centering
    \topcaption{
        Systematic uncertainties in percent in the ratio $R_{\Xbs}$ from the different sources
        and the total uncertainty, separately for the \jpsiXim and \jpsiLaKm decay modes.
        }
        \begin{scotch}{lcc}
            Source 
            &  \jpsiXim (\%)
            &  \jpsiLaKm (\%)
            \\  
            \hline
            \Xb fit model
            &   4.0
            &   6.9
            \\  
            \Xbs fit model
            &   7.7
            &   6.7
            \\ 
            Tracking efficiency
            &   5.2
            &   5.2
            \\  
            MC finite size
            &   6.5
            &   4.4
            \\[\cmsTabSkip]
            Total
            &   12.0
            &   11.8
        \end{scotch}
        \label{tab:syst_R_Xbs}
\end{table}

\subsection{Systematic uncertainties in the \texorpdfstring{\Xbs}{Xib*0} baryon mass and width measurements}

Several sources of systematic uncertainty are considered in the simultaneous measurement
of the \Xbs baryon mass difference and natural width.
To evaluate the systematic uncertainties related to the choice of functions used to fit
the \Xbs \dmX distributions, alternative functions are chosen and the maximum changes 
in the results of the fit are used to estimate the corresponding systematic uncertainty. 
We use a Student's t-distribution~\cite{Studentt} 
as the alternative function to describe the invariant mass resolution,
with the shape parameters determined from MC simulation. 
Fitting the data distributions leads to estimates for the systematic uncertainty of
$\pm0.003\MeV$ in the mass difference, while the change in the natural width is negligible.
 
We also vary the function used to describe the background in the fit.
We use the threshold function described earlier, multiplied by a first-order polynomial,
except for the \XbPX decay channel, where the number of events is too small to allow 
a reasonable fit to the background for functions with more parameters. 
Another alternative model uses the baseline background model to fit the same-sign $\Xb\pim$ 
distributions. The $\alpha$ values obtained in these fits are then used as fixed parameters
of the simultaneous fit.
From this, we estimate systematic uncertainties from this source of
$\pm0.002$ and $\pm0.04\MeV$ in the mass difference and natural width, respectively.

The systematic uncertainty coming from the choice of the fit range is estimated by varying
the \dmX fit region from $[0,\,0.05]$ to $[0,\,0.09]\GeV$. The maximum deviation of the fit
parameters is used as the systematic uncertainty, giving $\pm0.023$ and $\pm0.13\MeV$ in the
mass difference and natural width, respectively.

The signal shape for the \Xbs \dmX distribution is fit with a Gaussian resolution function,
convolved with a relativistic Breit--Wigner (RBW) and a Blatt--Weisskopf barrier 
factor~\cite{Blatt}, with the radial parameter in these functions set
to  $r = 3.5\GeV^{-1}$ and the angular momentum (spin) to $\ell = 1$.
To determine the systematic uncertainty associated with these choices, the fit is repeated
with the value of $r$ varied in the range 1--5 $\GeV^{-1}$ and $\ell$ set to 0 or 2.
The change in $r$ has a negligible effect, while the spin change leads to systematic uncertainties 
of $\pm0.022$ and $\pm0.02\MeV$ in the mass difference and natural width measurements, respectively.

For the \XbJX channel, we verify that the mass resolutions obtained in data and simulation
agree to within the combined uncertainty of 7.5\%.
We obtain a systematic uncertainty associated with any potential disagreement
in the \dmX mass resolution between data and simulation 
by repeating the \Xbs fit with the resolutions from MC scaled up or down by 1.075.
The resulting systematic uncertainties are $\pm0.004$ and $\pm0.08\MeV$
for the mass difference and natural width, respectively.

Simulation studies show a shift of $0.041\MeV$ between the reconstructed and generated mass differences for the \Xbs baryon peak;
this shift is treated as an additional systematic uncertainty in the \dmXfit measurement.

The systematic uncertainties described above are summarized in Table~\ref{tab:syst_Xbs},
together with the total systematic uncertainties, 
found from the sum in quadrature of those from the individual sources.

\begin{table}[ht]
    \centering
    \topcaption{
        The systematic uncertainties in \MeVns in the measurement of the \Xbs mass difference 
        and natural width from each of the sources, along with the total uncertainties.
    }
    \begin{scotch}{lcc}
        Source 
        &   \dmX (\MeVns)
        &   $\Gamma(\Xbs)$ (\MeVns)
        \\  
        \hline
        Signal model
        &   0.003
        &   $<0.01$
        \\  
        Background model
        &   0.002
        &   0.04
        \\  
        Fit range
        &   0.023
        &   0.13
        \\  
        RBW shape
        &   0.022
        &   0.02
        \\
        Mass resolution
        &   0.004
        &   0.08 
        \\
        Shift from reconstruction
        &   0.041
        &   \NA           
        \\[\cmsTabSkip]
        Total
        &   0.052
        &   0.16
    \end{scotch}
    \label{tab:syst_Xbs}
\end{table}

\section{Results
    \label{sec:results}}

Our final result for the ratio of the branching fractions for the \XbPX decay
with respect to the \XbJX normalization mode is
\begin{equation*}\begin{aligned}
\ifthenelse{\boolean{cms@external}}
{ 
    R & =  \dfrac{\br (\XbPX) }{\br (\XbJX) }\\
      & = 0.84^{+0.21}_{-0.19}\stat \pm 0.10\syst \pm 0.02\ebran, 
} 
{ 
    R =  \dfrac{\br (\XbPX) }{\br (\XbJX) } = 0.84^{+0.21}_{-0.19}\stat \pm 0.10\syst \pm 0.02\ebran, 
} 
\end{aligned}\end{equation*}
where the uncertainties are statistical, systematic, and related to the uncertainties
in the \jpsi and \psit branching fractions, respectively.
For the last term of Eq.~\eqref{eq:R}, we used the lepton universality assumption of 
$\br(\JMM) / \br(\PMM) = \br(\jpsi\to\Pep\Pem) / \br(\psit\to\Pep\Pem) = 7.53 \pm 0.17$ from the PDG~\cite{PDG}, 
since the dielectron modes are measured more precisely than the dimuon ones. 

Including the systematic uncertainties described in the previous section, 
the \Xbs mass difference and natural width are found to be 
\begin{equation*}\begin{aligned}
\ifthenelse{\boolean{cms@external}}
{ 
     M(\Xbs) &- M(\Xb) - m^{\mathrm{PDG}}(\pipm)\\
    & = 15.810 \pm 0.077 \stat \pm 0.052\syst\MeV, \\
    \Gamma(\Xbs) &= 0.87^{+0.22}_{-0.20} \stat \pm 0.16\syst\MeV.
} 
{ 
    M(\Xbs) - M(\Xb) - m^{\mathrm{PDG}}(\pipm) &= 15.810 \pm 0.077 \stat \pm 0.052\syst\MeV, \\
    \Gamma(\Xbs) &= 0.87^{+0.22}_{-0.20} \stat \pm 0.16\syst\MeV.
} 
\end{aligned}\end{equation*}
Using the world-average \Xb baryon mass~\cite{PDG}, our \dmXfit value corresponds 
to a \Xbs mass of $5952.4 \pm 0.1\,\text{(stat+syst)} \pm 0.6 \, (m_{\Xb})\MeV$, 
where the first uncertainty includes the statistical and systematic
components and the last comes from the uncertainty in the \Xb mass.
These measurements of the \Xbs baryon mass and width are significantly more precise
than the previous CMS results~\cite{BPH12001} 
and in agreement with those obtained by the LHCb experiment~\cite{LHCb_XibStar0, LHCb_Xbss}. 
Their recent measurement, using data corresponding to an integrated luminosity of 9\fbinv,
reported $\dmX = 15.80 \pm 0.02 \pm 0.01\MeV$ and $\Gamma(\Xbs) = 0.87\pm0.06\pm0.05\MeV$~\cite{LHCb_Xbss}. 

Finally, our measurement of the inclusive ratio of the \Xbs and \Xb production cross sections
gives
\begin{equation*}\begin{aligned}
\ifthenelse{\boolean{cms@external}}
{ 
    R_{\Xbs} & = \dfrac{\sigma(\pp\to\Xbs X) \, \br(\Xbsd)}{\sigma(\pp\to\Xb X)}\\
    & = 0.22 \pm 0.02 \stat \pm 0.02\syst,
} 
{ 
    R_{\Xbs} = \dfrac{\sigma(\pp\to\Xbs X) \, \br(\Xbsd)}{\sigma(\pp\to\Xb X)} = 0.22 \pm 0.02 \stat \pm 0.02\syst,
} 
\end{aligned}\end{equation*}
where we used the BLUE procedure~\cite{Lyons_combine, Blue_2, Blue_1} to combine the results
from the \XbJX and \XbJLK decay modes. 
The statistical and systematic uncertainties are assumed to be uncorrelated,
except for the tracking efficiency, which we treat as correlated.

\section{Summary and conclusions
    \label{sec:summary}}

In this article, we present the first observation of the \XbPX decay. 
We use data from  LHC proton-proton (pp) collisions at $\sqrt{s}= 13\TeV$, 
collected by the CMS experiment during 2016--2018, 
corresponding to an integrated luminosity of 140\fbinv.
We measure the ratio of the branching fraction for the new decay to that for \XbJX to be
\begin{equation*}\begin{aligned}
\ifthenelse{\boolean{cms@external}}
{ 
    R & =  \dfrac{\br (\XbPX) }{\br (\XbJX) }\\
    & = 0.84^{+0.21}_{-0.19}\stat \pm 0.10\syst \pm 0.02\ebran, 
} 
{ 
    R =  \dfrac{\br (\XbPX) }{\br (\XbJX) } = 0.84^{+0.21}_{-0.19}\stat \pm 0.10\syst \pm 0.02\ebran, 
} 
\end{aligned}\end{equation*}
where the last uncertainty comes from the uncertainties in the \jpsi and \psit branching fractions.

This result is consistent with analogous measured ratios from $\PB_{(\PQs)}$ and \Lb
decays such as $\Bp\to\PGy\Kp$, $\Bd\to\PGy\KS$, $\PBzs\to\PGy\phi$,
and $\Lb\to\PGy\La$, whose values are in the range 0.5--0.6~\cite{PDG} (here \PGy refers to the \jpsi and \psit mesons).
In general, currently existing results for such ratios do not form any clear and unambiguous 
pattern. New measurements, such as the one reported here, 
and corresponding theoretical predictions are required to build a robust model that can
reliably describe \PQb hadron decays to charmonium states.

We reconstruct \Xbs candidates using the \Xbsd decay mode 
by combining tracks from the \pp collision vertex 
with \Xb candidates from four different decay modes. 
A simultaneous fit of all decay modes is used
to extract the mass difference and natural width, which are 
consistent with our previous results~\cite{BPH12001}, but with much better precision.
They are also in agreement with the LHCb measurements~\cite{LHCb_XibStar0, LHCb_Xbss}.
Using the world-average value for the \Xb baryon mass~\cite{PDG}, 
we measure the mass of the \Xbs baryon to be
\begin{equation*}
M(\Xbs) = 5952.4 \pm 0.1\,\text{(stat+syst)} \pm 0.6 (m_{\Xb})\MeV,
\end{equation*} 
where the last uncertainty comes from the uncertainty in the \Xb baryon mass.
We measure the natural width to be $\Gamma(\Xbs) = 0.87^{+0.22}_{-0.20}\stat \pm 0.16\syst\MeV$.

{\tolerance=800
Finally, our determination of the \Xbs / \Xb relative production rate
$R_{\Xbs} = 0.22\pm0.02\stat\pm0.02\syst$
is in good agreement with the LHCb result~\cite{LHCb_XibStar0} 
of $0.28\pm0.03\pm0.01$ and is of a similar precision.
From the measured values of this ratio, we conclude that about
1/4 of \Xb baryons are produced from the \Xbsd decay.  
The other major \Xbs decay is $\Xbs\to\Xbz\piz$.
Since $\br(\Xbsns\to\Xbns\pion)$ should be close to 100\%, 
we expect $\br(\Xbsd) \approx 2 \, \br(\Xbs\to\Xbz\piz) \approx 2/3$, 
where the factor of 2 comes from isospin differences and the Clebsch--Gordan coefficients~\cite{PDG}.
Incorporating this estimate of $\br(\Xbsd)$ into our results for the ratio of production cross sections, we find that $\sigma(\pp\to\Xbs X) / \sigma(\pp\to\Xb X) \approx 1/3$. 
If the relative production rate for \Xbsm to \Xb follows the same scheme, 
the corresponding ratio can be estimated as 
$R_{\Xbsm} = [\sigma(\pp\to\Xbsm X) \mathcal{B}(\Xbsm\to\Xb\piz)] / \sigma(\pp\to\Xb X) \approx 1/3 \times 1/3 = 1/9 $.
Thus, we can conclude that about a third of the \Xb baryons are produced from \Xbsns decays.
\par}

Since decays from higher-mass excited \Xbns baryons are also possible, such as the $\Xbns(6227)$ doublet reported by the LHCb experiment~\cite{LHCb_Xib6227, LHCb_Xib6227_2020},
less than two thirds of the \Xb baryons are expected to be directly produced
from \pp collisions.
It is clear that further studies of different ground- and excited-state \Xbns baryons 
are needed to fully understand this family of baryons. 

\begin{acknowledgments}
We congratulate our colleagues in the CERN accelerator departments for the excellent performance of the LHC and thank the technical and administrative staffs at CERN and at other CMS institutes for their contributions to the success of the CMS effort. In addition, we gratefully acknowledge the computing centers and personnel of the Worldwide LHC Computing Grid and other centers for delivering so effectively the computing infrastructure essential to our analyses. Finally, we acknowledge the enduring support for the construction and operation of the LHC, the CMS detector, and the supporting computing infrastructure provided by the following funding agencies: SC (Armenia), BMBWF and FWF (Austria); FNRS and FWO (Belgium); CNPq, CAPES, FAPERJ, FAPERGS, and FAPESP (Brazil); MES and BNSF (Bulgaria); CERN; CAS, MoST, and NSFC (China); MINCIENCIAS (Colombia); MSES and CSF (Croatia); RIF (Cyprus); SENESCYT (Ecuador); ERC PRG, RVTT3 and MoER TK202 (Estonia); Academy of Finland, MEC, and HIP (Finland); CEA and CNRS/IN2P3 (France); SRNSF (Georgia); BMBF, DFG, and HGF (Germany); GSRI (Greece); NKFIH (Hungary); DAE and DST (India); IPM (Iran); SFI (Ireland); INFN (Italy); MSIP and NRF (Republic of Korea); MES (Latvia); LMTLT (Lithuania); MOE and UM (Malaysia); BUAP, CINVESTAV, CONACYT, LNS, SEP, and UASLP-FAI (Mexico); MOS (Montenegro); MBIE (New Zealand); PAEC (Pakistan); MES and NSC (Poland); FCT (Portugal); MESTD (Serbia); MCIN/AEI and PCTI (Spain); MOSTR (Sri Lanka); Swiss Funding Agencies (Switzerland); MST (Taipei); MHESI and NSTDA (Thailand); TUBITAK and TENMAK (Turkey); NASU (Ukraine); STFC (United Kingdom); DOE and NSF (USA).

\hyphenation{Rachada-pisek} Individuals have received support from the Marie-Curie program and the European Research Council and Horizon 2020 Grant, contract Nos.\ 675440, 724704, 752730, 758316, 765710, 824093, 101115353, and COST Action CA16108 (European Union); the Leventis Foundation; the Alfred P.\ Sloan Foundation; the Alexander von Humboldt Foundation; the Science Committee, project no. 22rl-037 (Armenia); the Belgian Federal Science Policy Office; the Fonds pour la Formation \`a la Recherche dans l'Industrie et dans l'Agriculture (FRIA-Belgium); the Agentschap voor Innovatie door Wetenschap en Technologie (IWT-Belgium); the F.R.S.-FNRS and FWO (Belgium) under the ``Excellence of Science -- EOS" -- be.h project n.\ 30820817; the Beijing Municipal Science \& Technology Commission, No. Z191100007219010 and Fundamental Research Funds for the Central Universities (China); the Ministry of Education, Youth and Sports (MEYS) of the Czech Republic; the Shota Rustaveli National Science Foundation, grant FR-22-985 (Georgia); the Deutsche Forschungsgemeinschaft (DFG), under Germany's Excellence Strategy -- EXC 2121 ``Quantum Universe" -- 390833306, and under project number 400140256 - GRK2497; the Hellenic Foundation for Research and Innovation (HFRI), Project Number 2288 (Greece); the Hungarian Academy of Sciences, the New National Excellence Program - \'UNKP, the NKFIH research grants K 124845, K 124850, K 128713, K 128786, K 129058, K 131991, K 133046, K 138136, K 143460, K 143477, 2020-2.2.1-ED-2021-00181, and TKP2021-NKTA-64 (Hungary); the Council of Science and Industrial Research, India; ICSC -- National Research Center for High Performance Computing, Big Data and Quantum Computing, funded by the NextGenerationEU program (Italy); the Latvian Council of Science; the Ministry of Education and Science, project no. 2022/WK/14, and the National Science Center, contracts Opus 2021/41/B/ST2/01369 and 2021/43/B/ST2/01552 (Poland); the Funda\c{c}\~ao para a Ci\^encia e a Tecnologia, grant CEECIND/01334/2018 (Portugal); the National Priorities Research Program by Qatar National Research Fund; MCIN/AEI/10.13039/501100011033, ERDF ``a way of making Europe", and the Programa Estatal de Fomento de la Investigaci{\'o}n Cient{\'i}fica y T{\'e}cnica de Excelencia Mar\'{\i}a de Maeztu, grant MDM-2017-0765 and Programa Severo Ochoa del Principado de Asturias (Spain); the Chulalongkorn Academic into Its 2nd Century Project Advancement Project, and the National Science, Research and Innovation Fund via the Program Management Unit for Human Resources \& Institutional Development, Research and Innovation, grant B37G660013 (Thailand); the Kavli Foundation; the Nvidia Corporation; the SuperMicro Corporation; the Welch Foundation, contract C-1845; and the Weston Havens Foundation (USA).
\end{acknowledgments}

\bibliography{auto_generated}
\cleardoublepage \appendix\section{The CMS Collaboration \label{app:collab}}\begin{sloppypar}\hyphenpenalty=5000\widowpenalty=500\clubpenalty=5000
\cmsinstitute{Yerevan Physics Institute, Yerevan, Armenia}
{\tolerance=6000
A.~Hayrapetyan, A.~Tumasyan\cmsAuthorMark{1}\cmsorcid{0009-0000-0684-6742}
\par}
\cmsinstitute{Institut f\"{u}r Hochenergiephysik, Vienna, Austria}
{\tolerance=6000
W.~Adam\cmsorcid{0000-0001-9099-4341}, J.W.~Andrejkovic, T.~Bergauer\cmsorcid{0000-0002-5786-0293}, S.~Chatterjee\cmsorcid{0000-0003-2660-0349}, K.~Damanakis\cmsorcid{0000-0001-5389-2872}, M.~Dragicevic\cmsorcid{0000-0003-1967-6783}, P.S.~Hussain\cmsorcid{0000-0002-4825-5278}, M.~Jeitler\cmsAuthorMark{2}\cmsorcid{0000-0002-5141-9560}, N.~Krammer\cmsorcid{0000-0002-0548-0985}, A.~Li\cmsorcid{0000-0002-4547-116X}, D.~Liko\cmsorcid{0000-0002-3380-473X}, I.~Mikulec\cmsorcid{0000-0003-0385-2746}, J.~Schieck\cmsAuthorMark{2}\cmsorcid{0000-0002-1058-8093}, R.~Sch\"{o}fbeck\cmsorcid{0000-0002-2332-8784}, D.~Schwarz\cmsorcid{0000-0002-3821-7331}, M.~Sonawane\cmsorcid{0000-0003-0510-7010}, S.~Templ\cmsorcid{0000-0003-3137-5692}, W.~Waltenberger\cmsorcid{0000-0002-6215-7228}, C.-E.~Wulz\cmsAuthorMark{2}\cmsorcid{0000-0001-9226-5812}
% ...
\par}
\vskip\cmsinstskip
\dag:~Deceased\\
$^{1}$Also at Yerevan State University, Yerevan, Armenia\\
$^{2}$Also at TU Wien, Vienna, Austria\\
% ...
\end{sloppypar}
\end{document}